\documentclass[a4paper,fleqn,usenatbib]{mnras}

\usepackage{graphicx,multicol}
\usepackage{times}

\usepackage{amssymb}   

\usepackage{newtxtext,newtxmath}

\usepackage[utf8]{inputenc}
\usepackage[T1]{fontenc}
\DeclareRobustCommand{\VAN}[3]{#2}
\let\VANthebibliography\thebibliography
\def\thebibliography{\DeclareRobustCommand{\VAN}[3]{##3}\VANthebibliography}

\usepackage{graphicx}	
\usepackage{amsmath}	
\usepackage{ulem}

\renewcommand{\v}[1]{\ensuremath{\mathbf{#1}}} 

\usepackage{xcolor}
\definecolor{blazeorange}{rgb}{1.0, 0.4, 0.0}
\definecolor{seagreen}{rgb}{0.18, 0.55, 0.34}
\definecolor{rufous}{rgb}{0.66, 0.11, 0.03}
\definecolor{royalfuchsia}{rgb}{0.79, 0.17, 0.57}
\definecolor{scarlet}{rgb}{1.0, 0.13, 0.0}
\definecolor{royalpurple}{rgb}{0.47, 0.32, 0.66}
\definecolor{darkblue}{rgb}{0, 0, 0.66}

\title[Insights on Prompt GRB Emission from Internal Shocks]{Prompt Gamma-Ray Burst Emission from Internal Shocks -- New Insights}

\author[Sk. Minhajur Rahaman]{
Sk. Minhajur Rahaman,$^{1}$\thanks{E-mail: rahaman.minhajur93@gmail.com}
Jonathan Granot,$^{1,2,3}$
Paz Beniamini$^{1,2,3}$
\\
$^{1}$Astrophysics Research Center of the Open University (ARCO), The Open University of Israel, P.O Box 808, Ra\'anana 4353701, Israel\\
$^{2}$Department of Natural Sciences, The Open University of Israel, P.O Box 808, Ra’anana 4353701, Israel \\
$^{3}$Department of Physics, The George Washington University, 725 21st Street NW, Washington, DC 20052, USA
}

\date{Accepted XXX. Received YYY; in original form ZZZ}

\begin{document}
\label{firstpage}
\pagerange{\pageref{firstpage}--\pageref{lastpage}}
\maketitle

\begin{abstract}
Internal shocks are a leading candidate for the dissipation mechanism that powers the prompt $\gamma$-ray emission in gamma-ray bursts (GRBs). In this scenario a compact central source produces an ultra-relativistic outflow with varying speeds, causing faster parts or shells to collide with slower ones. Each collision produces a pair of shocks -- a forward shock (FS) propagating into the slower leading shell and a reverse shock (RS) propagating into the faster trailing shell. The RS's lab-frame speed is always smaller, while the RS is typically stronger than the FS, leading to different conditions in the two shocked regions that both contribute to the observed emission. We show that optically-thin synchrotron emission from both (weaker FS + stronger RS)
can naturally explain key features of prompt
GRB emission such as the pulse shapes, time-evolution of the $\nu{}F_\nu$ peak flux and photon-energy, and the spectrum. Particularly,  it can account for two  features commonly observed in GRB spectra: (i) a sub-dominant low-energy spectral component (often interpreted as ``photospheric''-like), or (ii) a doubly-broken power-law spectrum with the low-energy spectral slope approaching the slow cooling limit. Both features can be obtained while maintaining high overall radiative efficiency without any fine-tuning of the physical conditions. 
\end{abstract}




\section{Introduction}

Internal shocks are invoked in a variety of astrophysical transients.  They are a  leading model for internal energy dissipation  in the prompt-emission phase of Gamma-Ray Bursts  
\citep[GRBs;][]{Rees1994,Sari1997,kobayashi1997,1998MNRAS.296..275D}. In the prompt GRB internal shocks model, a compact central source launches an ultra-relativistic outflow of  plasma with a varying velocity. At some distance from the source the faster  parts of the outflow overtake and collide with the slower parts of the outflow.

Few studies \cite[e.g.,][]{1998MNRAS.296..275D,2000ApJ...537..824S,2001ApJ...557..399G,2009A&A...498..677B,2017ApJ...837...33B, 2022MNRAS.511.5823R,2023ApJ...950...28R} have constructed prompt GRB lightcurves adopting a ``ballistic approach'' rather than solving for the hydrodynamic equations of shock propagation. In this approach  colliding shells are divided into discrete infinitely thin elements, a plastic collision of pairs of these discrete elements gives rise to a merged shell which again collides with another element and the process repeats. The internal energy dissipated in each collision of discrete elements is assigned to a forward shock if the Lorentz factor (hereafter LF, $\Gamma$) of the merged shell is closer to the slowly moving shell and vice versa. While this provides useful approximation for the LF of the shocked material and internal energy dissipation, this approach only provides a crude estimate for the location of the shock fronts. As we will show it is the relative location of the shock fronts and the different shock strengths that dictate the diversity of the light curves and the shape of the spectra. The ballistic approach washes away these features (see Appendix \ref{appD} for comparison of the lightcurves for the ballistic and our hydrodynamic approach).

\cite{2009MNRAS.399.1328G} parameterized the propagation of a single shock front assuming the same LF as the shocked matter, and emission between radii $R_\mathrm{o}$ and $R_\mathrm{f}$ \citep[building upon][]{Granot05,Gramot+08}. They found an analytic solution for the observed emission (lightcurves and spectra), using integration over the equal arrival time surface for a Band function emission spectrum. That work forms the foundation of this Letter. We make the following refinements: (i) we consider both shock fronts and estimate their LFs, $R_\mathrm{o}$ and $R_\mathrm{f}$ using the central source parameters, (ii) we account for the different LFs of each shock front and its shocked emitting matter.

To achieve efficient energy dissipation in internal shocks, the magnetization of the colliding shells cannot be too high. Moreover, the magnetization of the two colliding shells is typically expected to be comparable, as they are part of the same outflow (unlike the external forward and reverse shocks that form as the ejecta is decelerated by the external medium). A useful approach is to consider a single collision where the faster (trailing)  and slower (leading) parts of the outflow are approximated as two discrete uniform cold shells. Post collision, a contact discontinuity (CD) forms between the shells and a pair of shocks is launched. The slower (leading) shell is shocked by a forward shock (FS) while the faster (trailing) shell is shocked by the reverse shock (RS). The shocks dissipate part of the outflow's kinetic energy into internal energy. The RS front’s lab-frame speed is always
smaller than that of the FS front. Moreover, the RS is typically stronger than the FS, which leads to different conditions in the two shocked
regions, both of which contribute to the observed emission. 

The present study is the first prompt GRB internal shocks modeling to self-consistently account for both shock fronts. In our treatment, we solve the hydrodynamics for shock propagation of both shocks and supplant it by calculating the observed optically thin synchrotron emission through integration over the equal arrival time surface for each shock, and adding these two contributions. 

This letter is structured as follows. \S\,\ref{set-up} describes the hydrodynamical setup for shock propagation, the particulars of the synchrotron emission process and the calculation of the observed radiation. \S\,\ref{Results} describes the lightcurves (pulse morphology), the temporal  evolution of the instantaneous spectra and  properties of the time-integrated spectrum. \S\,\ref{summary} summarizes our key findings.

\section{the basic setup of our model} \label{set-up}

 Here we describe the basic setup of (i) the shock hydrodynamics in the lab frame (\S\,\ref{hydro}), (ii) the prescription for the underlying emission mechanism in the co-moving frame of the shocked fluid (\S\,\ref{sec:emission}), and (iii) details of the equal arrival time surfaces in the observer frame, used for calculating the observed radiation (\S\,\ref{EATS_sec}).

Our analysis employs the following three reference frames: (i) the lab frame associated with the central  source, (ii) the local comoving frame  of the shocked fluid, and (iii) the observer frame of an observer receiving the photons. Frames (i) and (iii) are essentially the same (up to cosmological corrections), but refer to the photon emission (or lab-frame) time $t$ and arrival time $T$ at the observer, respectively. Quantities in the co-moving frame are denoted with primes.

\subsection{Hydrodynamics of shock propagation}\label{hydro}

\cite{rahaman2023internal} (hereafter Paper I) provide an in-depth analysis of the 1D shock hydrodynamics post-collision. Here we summarize some key results for a collision of ultra-relativistic shells, relevant to our case. Observations suggest that the peak flux of the prompt GRB pulses from a given burst do not vary considerably. This provides a particularly good motivation for assuming a constant source power. Further, it is found that the duration of the prompt pulses are similar to the separation of the pulses \citep{Nakar2002}. This gives good motivation for similar shell ejection and source inactivity timescales. 

A central  source of constant (isotropic equivalent) power $L$  ejects two discrete cold and unmagnetized shells (S1,\,S4) over activity timescales $(t_\mathrm{on1},\,t_\mathrm{on4})$  separated by an inactivity time $t_\mathrm{off}$. The leading shell S1 and trailing shell S4 move with ultra-relativistic proper speeds ($u=\Gamma\beta$), $u_{4}>u_{1} \gg 1$ with $a_u=u_{4}/u_{1}>1$. The front and back edges of shells (S1,\,S4) are ejected at times $(t_\mathrm{ej,f1},\,t_\mathrm{ej,b1})$ and $(t_\mathrm{ej,f4},\,t_\mathrm{ej,b4})$,  corresponding to the ejection timescales: $t_\mathrm{on,i} = t_\mathrm{ej,bi} - t_\mathrm{ej,fi}$ for $i =\;$(RS,\,FS).  The front edge of shell S4 collides head-on with the back edge of shell S1 at a distance $R_\mathrm{o}$ from the central  source, and at time $t_\mathrm{o}$ where
\begin{equation}
R_\mathrm{o} = \frac{\beta_1 \beta_4 c t_\mathrm{off}}{(\beta_4 - \beta_1)} \approx \frac{2\Gamma_1^2ct_\mathrm{off}}{1-a_u^{-2}}\ ,\quad 
t_\mathrm{o} - t_\mathrm{ej,f4} = \frac{\beta_{1} t_\mathrm{off}}{(\beta_4 - \beta_1 )} \approx \frac{R_\mathrm{o}}{c}\ . 
\end{equation}

The collision produces a pair of shocks, where the two shocked parts of shells S1 and S4 are separated by a CD and move with the same proper speed $u$.  
The proper speeds of the FS (RS) propagating into shell S1 (S4) satisfy $u_{\rm RS}<u<u_{\rm FS}$. All three proper speeds remain constant in planer geometry (that we assume here for the dynamics, for simplicity). The shock fronts and CD divide the shells (S1,\,S4) into four regions (R1,\,R2,\,R3,\,R4). Regions R1/R2 and R4/R3 are the unshocked/shocked parts of shells S1 and S4, respectively. The requirement of equal pressure and velocity across the CD, implies an equal ram pressure
across the CD in its rest frame, $u_{21}^2=fu_{34}^2$ were $u_\mathrm{ij}^2=\Gamma_\mathrm{ij}^2-1$, 
$\Gamma_{21} = \Gamma_2 \Gamma_1 (1 - \beta_1 \beta_2)$ and $\Gamma_{34} = \Gamma_3 \Gamma_4 (1 - \beta_3 \beta_4)$. 
The shock strength (internal energy per unit rest energy in the shocked region), $\Gamma_\mathrm{ij} - 1$ where ij\;=\;(21,\,34), 
is governed by the proper density contrast, $f =  {n'_4}/{n'_1}$, which for a constant source power  is  $\sim a^{-2}_\mathrm{u} \ll 1$. 
This shows that the RS strength $(\Gamma_{34} -1)$ is expected to be larger than the FS strength $(\Gamma_{21}-1)$.
The proper velocity of the shocked fluid for an ultra-relativistic flow is $u \approx \Gamma =  [(\sqrt{f} a^2_\mathrm{u} + a_\mathrm{u})/(a_\mathrm{u} + \sqrt{f})]^{1/2}\,\Gamma_1$ (see Appendix \ref{appA}). The FS reaches the front edge of shell S1 in time $t_\mathrm{FS}$ while the RS reaches the back edge of shell S4 in time $t_\mathrm{RS}$. At the instant $(t_\mathrm{o}+t_\mathrm{FS},\, t_\mathrm{o} + t_\mathrm{RS})$ the final location of the FS and RS fronts is ($R_\mathrm{FS},\,R_\mathrm{RS}$) such that
\begin{equation}
    \frac{\Delta R_\mathrm{i}}{R_\mathrm{o}} = \frac{R_\mathrm{f,i} - R_\mathrm{o}}{R_\mathrm{o}} =\frac{\beta_i t_i}{R_\mathrm{o}/c}\approx\frac{(1-a_u^{-2})t_i}{2\Gamma_1^2t_{\rm off}} \quad \text{for $i=\;$ (RS,\,FS)}\ ,
\end{equation}
We account for spherical expansion with a hybrid approach, in which the proper speeds ($u$, $u_{\rm RS}$, $u_{\rm FS}$) remain constant 
($\propto R^{0}$) while the matter density 
varies as $\rho \propto R^{-2}$. The first assumption allows analytic solutions of the shock crossing times. The second assumption allows us to account for the decrease in density as the shells propagate outward.
The shell crossing timescales are given by (see Appendix C of Paper I),
\begin{equation}
      t_\mathrm{RS} = \frac{\beta_4 t_\mathrm{on4}}{\beta_4 - \beta_\mathrm{RS}} \approx\frac{2\Gamma^2t_\mathrm{on4}}{g_{\rm RS}^2-\left(\frac{\Gamma}{\Gamma_4}\right)^2}\ ,\quad 
 t_\mathrm{FS} =  \frac{ \beta_1 t_\mathrm{on1}}{\beta_\mathrm{FS} - \beta_1} \approx\frac{2\Gamma^2t_\mathrm{on1}}{\left(\frac{\Gamma}{\Gamma_1}\right)^2-g_{\rm FS}^2}\ ,
\end{equation}
where $g_{\rm RS}=\Gamma/\Gamma_{\rm RS}>1$ and $g_{\rm FS}=\Gamma/\Gamma_{\rm FS}<1$.
 For the rest of our analysis, we fix the proper speeds to $(u_{1},\,u_{4}) =(100,\,200)$ corresponding to a proper speed contrast of $a_\mathrm{u} = 2$ while our fiducial ratios of the activity and inactivity timescales are
 $(t_\mathrm{on1}:t_\mathrm{off}:t_\mathrm{on4}) = (1:1:1)$. Our fiducial case corresponds to the collision of two equal energy shells. We also fix $t_\mathrm{off}$ which in turn fixes the fiducial collision radius $R_\mathrm{o}$ and time $t_\mathrm{o}$. In \S\,\ref{Results} we will explore different combinations of the ratio $t_\mathrm{on1}:t_\mathrm{on4}$. We next discuss why even if this ratio is changed arbitrarily, it has  a relatively modest effect on the lightcurves. 

For fixed shell proper speeds $(u_{1},\,u_{4})$ and a constant source power, the shock front LFs do not change  when varying the ejection timescales, as the proper density contrast $f$ remains unaltered. For an ultra-relativistic flow, the radial width of each shell scales linearly with its ejection duration, $\Delta_\mathrm{i}=\beta_ict_{\textrm{on},i}\approx ct_{\textrm{on},i}$,
and therefore so does the shock crossing duration $t_\mathrm{i}$. However, a shock cannot always cross the whole shell, for the following reason.  The shell crossing timescales are generally unequal, $t_\mathrm{RS} \neq t_\mathrm{FS}$. For example, when $t_\mathrm{on1} = t_\mathrm{on4}$ the RS crosses first,
$t_\mathrm{RS} < t_\mathrm{FS}$. 
When the RS reaches the back edge of the trailing shell S4, the rear edge of the high-pressure region R3 faces vacuum. 
This creates a rarefaction (rf) wave that propagates from the vacuum interface into region R3, towards the CD and FS front. If $\Delta_1$ is large enough, the rf wave can catch up with the FS front and  stall it, suppressing further internal energy dissipation. Hence, only when $t_\mathrm{on1}/t_\mathrm{on4}\approx\Delta_1/\Delta_4$ is sufficiently close to unity can both shocks finish crossing their respective shells. For our setup, this corresponds to the limits $0.42 \leq {t_\mathrm{on1}}/{t_\mathrm{on4}} \leq 2.68$ (see Appendix H of Paper I), beyond which further varying this ratio stops affecting the emission. To respect this limit, for all the illustrations we will restrict the ratio to $0.5 \leq t_\mathrm{on1}/t_\mathrm{on4} \leq 2$. 

\subsection{Synchrotron emission process}\label{sec:emission} 

As shown in Appendix \ref{appD}, 
for our moderate fiducial 
proper speed contrast ($a_\mathrm{u} = 2$) the  energy dissipation efficiency (defined as the ratio of the internal energy produced
to the original total kinetic energy)  
associated with the RS and the FS is $\sim\,$5\% and 3\%, respectively. The overall efficiency of $\sim$ 9 \% is consistent with what has been estimated for plastic collision in internal shock models  \citep[][see also Appendix \ref{appD} for a comparison of the hydrodynamic and ballistic approaches]{Rees1994,kobayashi1997,Krimm07}. 
Moreover, model-independent constraints on the prompt efficiency from combined prompt and afterglow observations give similar estimates (see \citealt{Beniamini2015}).
 The observed radiation is from a population of shock-accelerated non-thermal electrons
that constitute  a fraction $\xi_\mathrm{e}$ of 
all electrons and carries a fraction $\epsilon_\mathrm{e}$ of the 
dissipated internal energy density, $e'_{\rm int}$. A fraction $\epsilon_\mathrm{B}$ of the dissipated internal energy is carried by the magnetic field, 
$B^{\prime\,2}/8\pi=\epsilon_Be'_{\rm int}$. Since only a fraction $\epsilon_{\gamma} = \epsilon_\mathrm{e}\epsilon_\mathrm{rad} $ of the dissipated energy can be radiated, in order to satisfy the observed energetics it is essential  that the radiation be close to the fast cooling regime ($\epsilon_\mathrm{rad} \approx 1$). Thus, the RS should be fast cooling or at least not very slow cooling.

To fulfill the energy requirements implied by prompt GRB observations, 
we assume the fast cooling regime of optically thin synchrotron emission from non-thermal power-law electrons, with a comoving energy distribution
$dn'/d\gamma_e \propto \gamma_e^{-p}$ for $\gamma_\mathrm{m} \leq \gamma_e \leq \gamma_\mathrm{M}$. The emission is assumed to be isotropic in the comoving frame of the shocked fluid.  For simplicity, we assume the microphysical parameters $(\epsilon_\mathrm{e},\epsilon_\mathrm{B},\xi_\mathrm{e},p)$ to be the same in both shocked regions. As the shock crossing times are similar, the comoving dynamical times 
are also similar and so is the cooling LF $\gamma_c = {6 \pi m_e c}/{\sigma_\mathrm{T} B'^2 t'_\mathrm{dyn}}$, 
where $m_\mathrm{e}$ ($m_\mathrm{p}$) is the electron (proton) mass, 
$c$ is the speed of light and $\sigma_\mathrm{T}$ is the Thomson 
cross-section. 
The minimum electron LF (for $p>2$) is given by $\gamma_\mathrm{m} = \frac{p-2}{p-1} \frac{m_\mathrm{p}}{m_\mathrm{e}}\frac{\epsilon_\mathrm{e}}{\xi_\mathrm{e}} (\Gamma_\mathrm{ij} -1) \propto \Gamma_\mathrm{ij} - 1$, where ij = (21,\,34), and is hence
higher for the RS compared to the FS. Therefore, we can have $\gamma_\mathrm{m} \sim \gamma_\mathrm{c}$ in the forward shocked region and $\gamma_\mathrm{m} > \gamma_\mathrm{c}$ in the reversed shocked region. This gives a natural motivation for choosing a marginally fast cooling ($\gamma_\mathrm{m} \sim \gamma_\mathrm{c}$) for the forward shocked region R2 and  fast cooling ($\gamma_\mathrm{m} \gg \gamma_\mathrm{c}$)  for the reverse shocked region R3. 
Hereafter we assume $\epsilon_{\rm rad}=0.5$ and $\epsilon_{\rm rad}=1$ for the marginally cooling FS and the fast cooling RS, respectively. In \S\,\ref{Results} we explore the effects of a fast cooling FS on the time-integrated spectrum.

\subsection{Equal arrival time surfaces} \label{EATS_sec}

\begin{figure}
\centering
\includegraphics[scale=0.5]{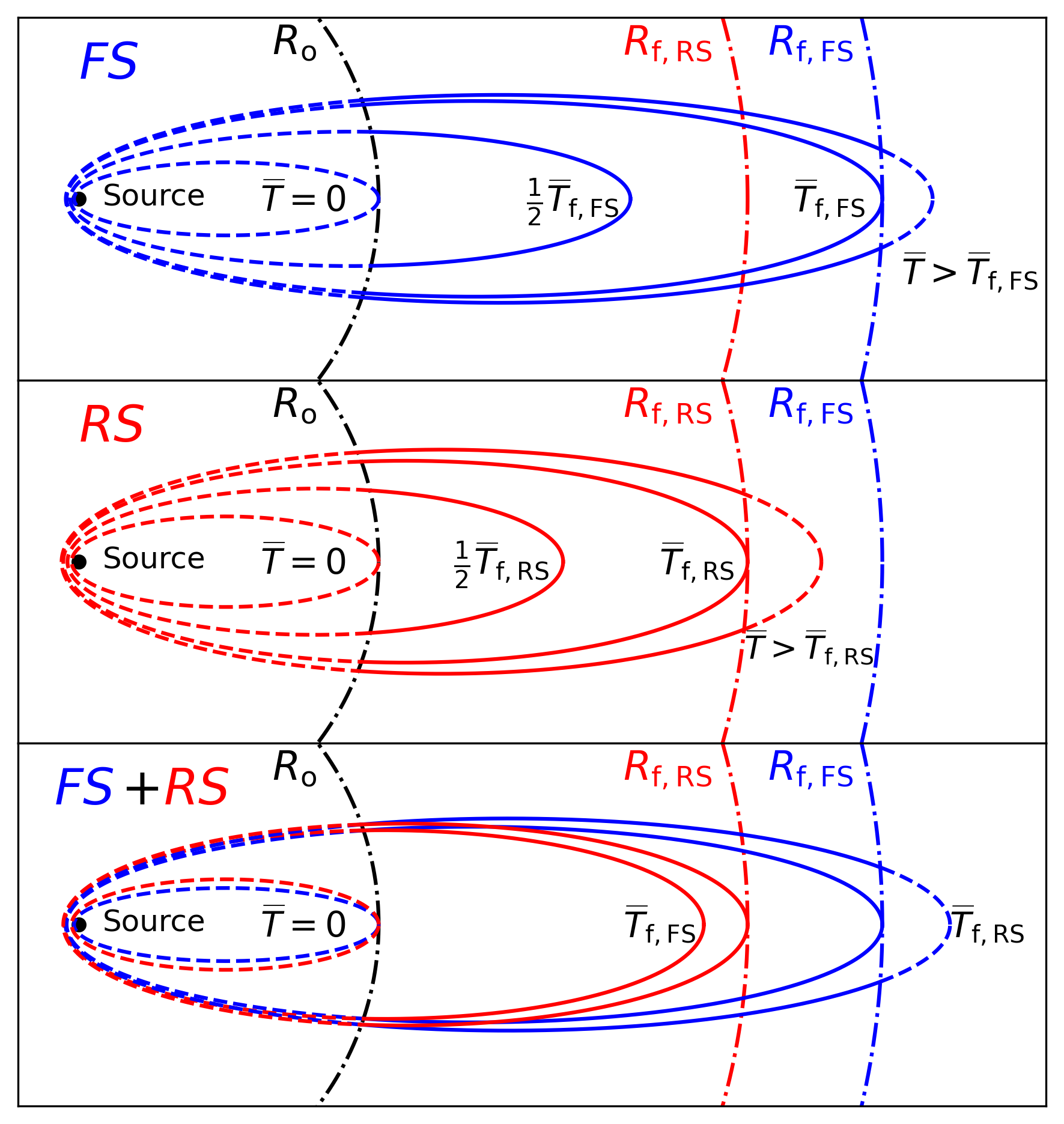} 
\caption{ Illustration of the equal arrival time surfaces (EATS) for the time combination $t_\mathrm{on1}:t_\mathrm{off}:t_\mathrm{on4} =(1:1:1)$. The prescription for construction of EATS is provided in Appendix \ref{appG}. While for the rest of the graphs we use values of $(u_{1},u_{4}) = (100,200)$, (only) here for illustrating the EATS we use 
instead $(u_{1},u_{4}) = (3,6)$  corresponding to $(u_\mathrm{FS},\,u_\mathrm{RS}) = (4.05,\,3.24)$. The EATS major to minor axis ratio is the shock front LF. The black dot corresponds to the compact central source. The observer is located far to the right.
The EATS for the forward shock (FS) and reverse shock (RS) are shown by blue and red lines, respectively. For a given EATS only the solid portion contributes to the observed emission. \textbf{\textit{Top $\&$ Middle panels}}: EATS for the FS $\&$ RS, each shown at four instants (see text for explanation).\textbf{\textit{Bottom panel}}: the combined EATS for both shock fronts (the outer ones belong to the FS while the inner ones belong to RS ) at the instants $\overline{T} = (0,\overline{T}_\mathrm{f,FS}, \overline{T}_\mathrm{f,RS})$.} 
\label{EATS_fig}
\end{figure}

The equal arrival time surface (EATS) for each shock front $i$ is defined as the locus of emission points from which photons 
reach an observer at a given  observed time, $T=t- R\cos\theta/c$. For a constant shock front LF $\Gamma_\mathrm{i}$ the EATS is an ellipsoid with a major to minor axis ratio of $\Gamma_\mathrm{i}$. The largest radius along the EATS (at $\theta=0$) is 
\begin{equation}
    R_\mathrm{L,i} =\frac{\beta_\mathrm{i}cT_{z,\textrm{i}}}{1-\beta_\mathrm{i}} 
    \approx 2 c \Gamma^2_\mathrm{i}T_{z,\textrm{i}}\ \ ,\quad\quad T_{z,\textrm{i}}\equiv \frac{T - T^\mathrm{eff}_\mathrm{ej,i}}{1+z}\ , 
\end{equation}
(see Eq. 9 of Appendix \ref{appH}), where $z$ is the redshift and $T^\mathrm{eff}_\mathrm{ej,i}$ is the arrival time of a hypothetical photon emitted at the source at the effective ejection time 
$t^\mathrm{eff}_\mathrm{ej,i}$ of a shell moving at $\beta_\mathrm{i}$ that coincides with the shock at $R\geq R_\mathrm{o}$
(see Appendix \ref{appG} and \ref{appH} for details).
The first photons from both shocks 
reach the observer
at $T_\mathrm{s}$, as both are emitted at $R_\mathrm{o}$, $t_\mathrm{o}$ and $\theta=0$.
However, as $\Gamma_{\rm FS}>\Gamma_{\rm RS}$, for our fiducial parameter values
the signal from the FS fully crossing S1 arrives before that from the RS fully crossing  S4, even though $t_{\rm RS}<t_{\rm FS}$.

Fig. \ref{EATS_fig} shows the EATS due to both shock fronts.  
We define a normalized time $\overline{T}_\mathrm{i} =(T-T_\mathrm{s})/T_\mathrm{0,i}$ where $T_\mathrm{0,i}=T_\mathrm{s}-T^\mathrm{eff}_\mathrm{ej,i}$
such that $\overline{T}_\mathrm{i} = 0$ 
is the time of the first photons from both shocks, emitted at $(R_\mathrm{L,o},\theta=0)$.
For display purposes we use $T_\mathrm{0,RS}$ as the normalization constant, i.e. use $\overline{T}=\overline{T}_{\rm RS}$;
$\overline{T} = \overline{T}_\mathrm{f,i}$ 
are the arrival time of photons emitted at $(R_\mathrm{L,f,i},\theta=0)$ 
(see Appendix \ref{appG} for details). The top and middle panels show the EATS due to both shock fronts at four instants. The times $\overline{T} = (0, 0.5 \overline{T}_\mathrm{f,i}, \overline{T}_\mathrm{f,i})$ correspond to 
$R_\mathrm{L,i}=R_\mathrm{o}$, $\frac{1}{2}(R_\mathrm{f,i} +R_\mathrm{o})$ and $R_\mathrm{f,i}$. 
At  $\overline{T} > \overline{T}_\mathrm{f,i}$ we have $R_\mathrm{L,i}>R_\mathrm{f,i}$ and the observed flux is dominated by high-latitude emission \citep[HLE;][]{PK2000}, where contributions to the observed flux come from $R_\mathrm{o}\leq R\leq R_\mathrm{f,i}$ corresponding to $\theta_{\rm min}\leq\theta\leq\theta_{\rm max}$ where $\theta_{\rm min}=[2c(T-T_\mathrm{f,i})/R_\mathrm{f,i}]^{1/2}$ and $\theta_{\rm max}=[2c(T-T_\mathrm{s})/R_\mathrm{o}]^{1/2}$. The bottom panel shows the EATS for both shock fronts at the same three normalized times.

\section{Results}\label{Results}

\begin{figure*}
\centering
\includegraphics[width=0.32\textwidth,height=0.333\textheight,{trim=7.5 15 8 6.9},clip]{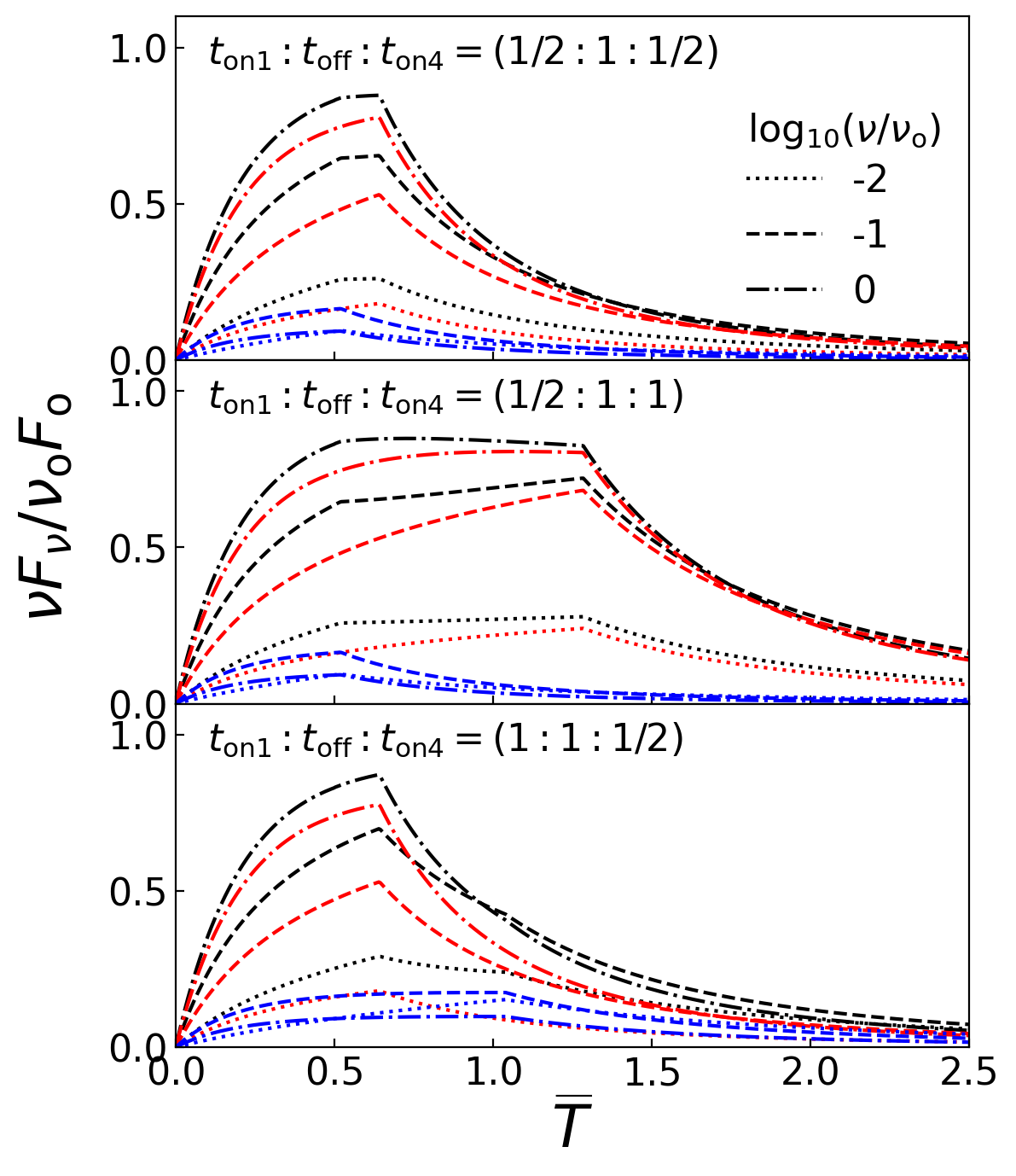}
\hspace{0.1cm}
\includegraphics[width=0.32\textwidth,height=0.333\textheight,{trim=7.5 12 7 7},clip]{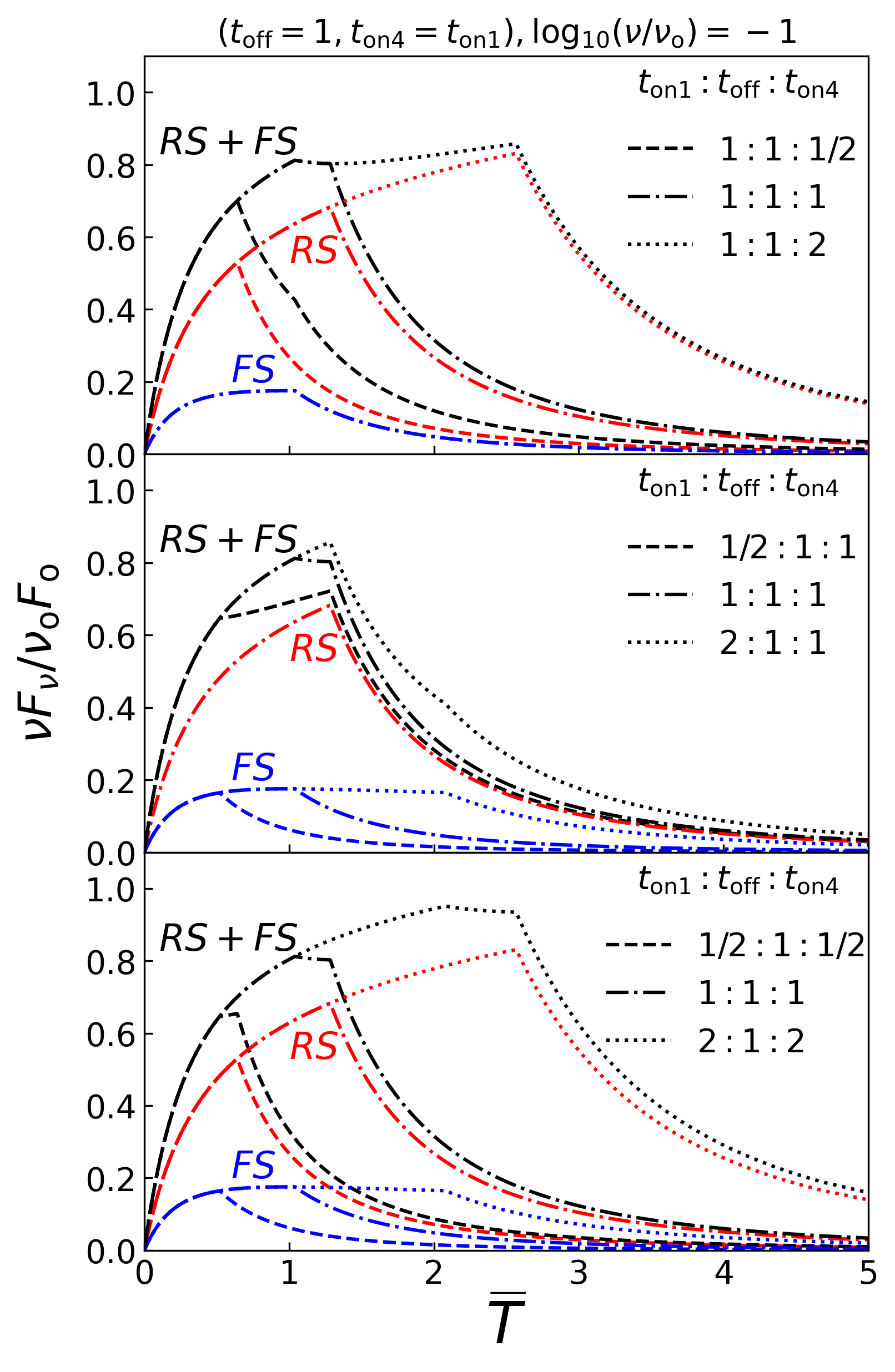}
\hspace{0.35cm}
\includegraphics[width=0.32\textwidth,height=0.333\textheight,{trim=7.5 8 8 7},clip]{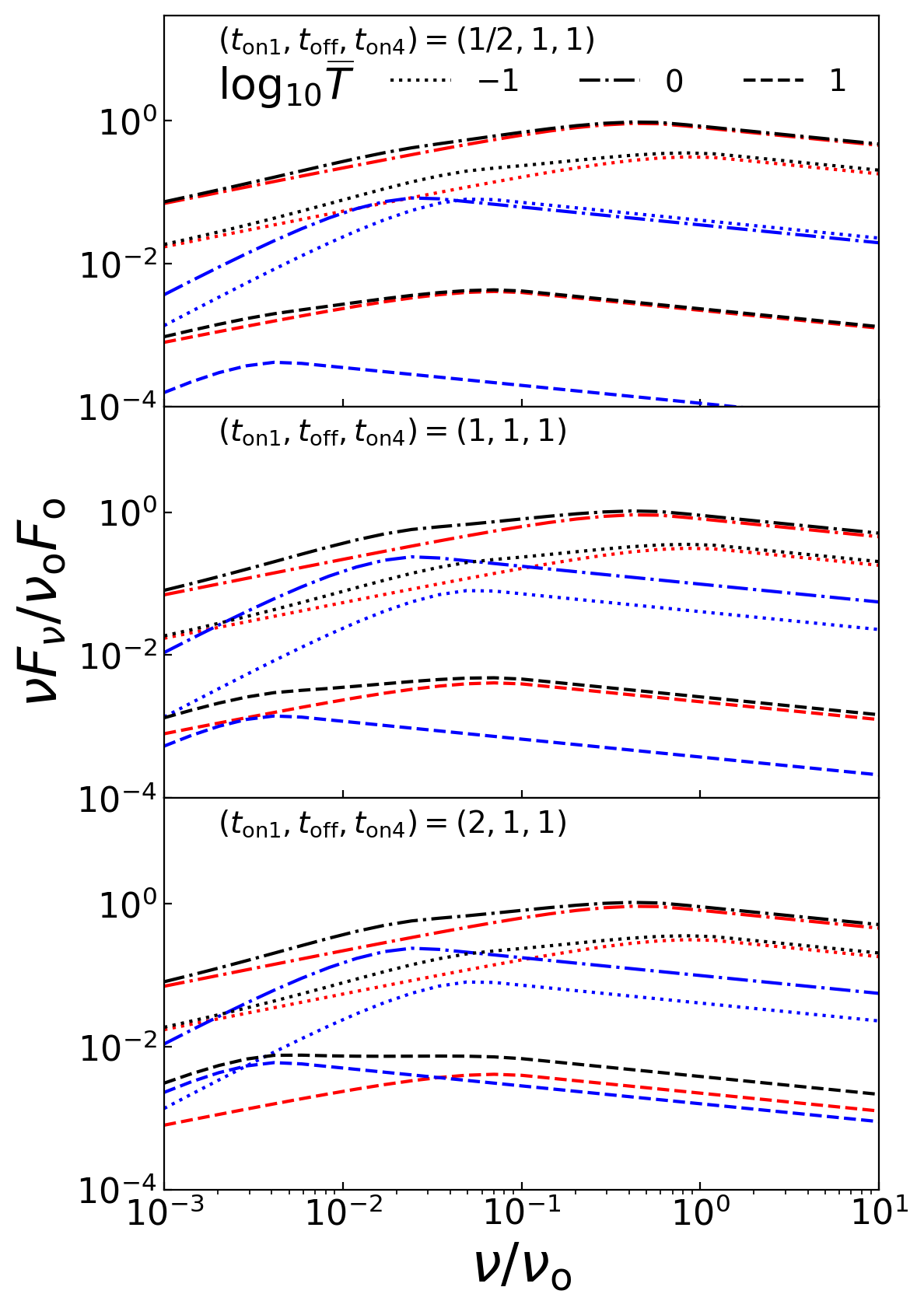}    
\caption{In all panels, red, blue and black lines correspond to the RS, the FS and their sum, respectively (see text for details). \textbf{\textit{Left}}: Lightcurves at different frequencies. 
\textbf{\textit{Middle}}: Lightcurves at a fixed frequency of $\log_{10}(\nu/\nu_\mathrm{o}) = -1$ and varying values of $t_{\rm on1}:t_{\rm off}:t_{\rm on4}$. \textbf{\textit{Right}}: Time resolved spectrum at three normalized times. The dotted line, the dot-dashed line and the dashed line correspond to early, intermediate and late times, respectively (see text for details).
}
\label{LC_spectra}
\end{figure*}

The purpose of this present section is to explore the pulse morphology and spectra of GRBs. To calculate
the observed flux at a given observed time we use
the formalism of \cite{2009MNRAS.399.1328G} and further refine it
by distinguishing between the shock front LF, $\Gamma_\mathrm{i}$, and the shocked matter LF, $\Gamma$ 
(common to both the shocked regions), requiring $g_\mathrm{i} \neq 1$ (see \S\,\ref{hydro} for definition). The flux density $F_{\nu}$ received by an on-axis observer from shock
$i$ at the normalized time $\widetilde{T}_{i}  = \overline{T}_\mathrm{i}+1$ is given by 
\begin{equation}\label{flux}
     \frac{F_{\nu}}{F_\mathrm{o,i}} (\widetilde{T}_\mathrm{i})= g^2_{i} \widetilde{T}_\mathrm{i} \int^{y_\mathrm{max}}_{y_\mathrm{min}} \; dy\;  \frac{y^{-1}}{[1+g^2_\mathrm{i}(y^{-1} - 1)]^3} \; S\left[ x \right]\ ,  
\end{equation}
where  $y = R/R_\mathrm{L,i}(\widetilde{T}_\mathrm{i})$ ($y_{\rm min},y_{\rm max}$  and the full derivation of the expression are described in Appendix \ref{appH}) where $R_\mathrm{L,i}$ is the largest radius on the EATS at the normalized time  $\widetilde{T}_\mathrm{i} = (T - T^\mathrm{eff}_\mathrm{ej,i})/T_\mathrm{0,i}$ (see Appendix \ref{appG}) ,  $F_\mathrm{o,i} 
= 2 \Gamma (1+z) L'_\mathrm{o,i}/4 \pi d^2_{L}$ where $d_\mathrm{L}$ is the luminosity distance, $z$ is the redshift and $L'_\mathrm{o,i}$ is the peak luminosity in the comoving frame at the collision radii,  $x=\nu'/\nu'_\mathrm{p}$ where $\nu'$ and $\nu'_\mathrm{p}$ are the comoving frequency and peak frequency respectively, and $S[x]$ is the normalized Band-function given by
\begin{equation}
  S[x] = e^{1+b_1} 
  \begin{cases}
      &\ x^{b_{1}} e^{-(1+b_1) x}  \hspace{2cm} \text{$x\leq x_\mathrm{b}$\ ,} \\
      &\ x^{b_{2}} x^{b_1 - b_2}_\mathrm{b} e^{-(b_1 - b_2)} \hspace{1.1cm}  \text{$x\geq x_\mathrm{b}$\ ,}
  \end{cases} \label{Sx}
\end{equation}
where $b_1$ and $b_2$ are the asymptotic low and high-frequency spectral slopes, which satisfy $b_1>-1>b_2$, and $x_\mathrm{b} = (b_1 - b_2)/(1+b_1)>1$. Appendix \ref{appH} gives analytical estimates for the flux from Eq.~(\ref{flux}). For both shocks, the following relationship holds (see Appendix \ref{appE} \& \ref{appF})
\begin{equation}
    \frac{\nu_\mathrm{o,FS}}{\nu_\mathrm{o,RS}} \approx \left( \frac{\Gamma_{21}\!-\!1}{\Gamma_{34}\!-\!1} \right)^{2}\,,\ \ \ \,  \frac{F_\mathrm{o,FS}}{F_\mathrm{o,RS}} \approx \left( \frac{\epsilon_\mathrm{rad,FS}}{\epsilon_\mathrm{rad,RS}} \right)\; \left( \frac{\Gamma_{34}\!-\!1}{\Gamma_{21}\!-\!1} \right)^{2} \left( \frac{\beta_{21}}{\beta_{34}} \right)\,. \label{freq_flux0}     
\end{equation}
The hydrodynamical and emission  parameter values for the two shocked regions are given
in Table~\ref{Param_space}. For purposes of display, all the flux, frequency, and observed times are normalized to those of the RS such that $F_\mathrm{o} \equiv F_\mathrm{o,RS} , \nu_\mathrm{o} \equiv \nu_\mathrm{o,RS}, T_\mathrm{0} \equiv T_\mathrm{0,RS}$ (see Appendix \ref{appI}). For all figures (except Fig. \ref{integ_spectra}), we assume a marginally fast cooling FS. The RS shock is always in the fast cooling regime.

\begin{table}
    \centering
    \caption{Parameter space (hydrodynamics and synchrotron emission) for both the shocked regions. The proper speeds $(u_1,u_4)$ of shells S1 and S4 are fixed to be 100 and 200 respectively. The indices $(b_1,b_2)$ correspond to the Band function defined in equation \ref{Sx}. The quantity $p$ is the power-law index of the non-thermal electrons and is taken to have the fixed value $2.5$.  For the FS the quantity $(b_1 = 1/3, \epsilon_{\rm rad} = 0.5)$ corresponds to the marginally fast cooling regime of synchrotron emission. For both shocks the combination $(\epsilon_{\rm rad},b_1,b_2) = (1,-1/2, -p/2)$ corresponds to the very fast cooling regime of synchrotron emission. The hydrodynamical quantities $g_\mathrm{i}$ and shock strength $\Gamma_\mathrm{ij} - 1$ are defined in \S\,\ref{hydro}.  }
    \begin{tabular}{cccccc} \hline 
     Shock front    & $\epsilon_{\rm rad}$ & $b_\mathrm{1}$ & $b_\mathrm{2}$ & $g_\mathrm{i}$ & $\Gamma_\mathrm{ij} - 1$ \\ \hline 
       FS           & 0.5 (1) & 1/3 (-1/2)  & -p/2 = - 1.25 & 0.926 & 0.027\\
       RS           &  1.0 & -1/2 & - p/2 = -1.25 &1.155 & 0.107 \\ \hline 
    \end{tabular}
    \label{Param_space}
\end{table}

The left panels of Fig.~\ref{LC_spectra} show lightcurves at different frequencies. Each panel displays a fixed time ratio
$t_\mathrm{on1}:t_\mathrm{off}:t_\mathrm{on4}$. 
The pulse from each shock peaks at 
$\overline{T}_\mathrm{f,i}$, and the subsequent 
tail is due to HLE. The total pulse has a particularly complex morphology
if $\overline{T}_\mathrm{f,RS}$ and $\overline{T}_\mathrm{f,FS}$ 
are well separated as occurs in middle and bottom panels.

The middle panels of Fig.~\ref{LC_spectra} show the lightcurves at a fixed frequency. In the top panel the FS pulse 
is fixed while the RS pulse is varied by increasing the ejection duration $t_\mathrm{on4}$ and thus the RS pulse width and peak time. As a result, the 
total pulse  shows 
a plateau at the limit $t_{\rm on4}\gg t_{\rm off}$. In the middle panel the RS pulse is fixed and the FS pulse is varied, producing comparatively narrow pulse widths 
and plateau regions in the observed profile. In the bottom panel both pulses due to FS and RS are varied and we obtain very large pulse widths and narrow plateaus in the observed profile.

The instantaneous spectrum due to shock front $i$ can be well-modeled by (with an error of less than 1\%)
\begin{eqnarray}\label{inst_spectra1}
    (\nu F_{\nu})_\mathrm{i} &=& \nu_\mathrm{pk,i} F_\mathrm{\nu_\mathrm{pk,i}} \; S\left[ {\nu}/{\nu_\mathrm{pk,i}} \right]\ , 
    \\ \label{inst_spectra2}
     \frac{\nu_\mathrm{pk,i} F_{\nu_\mathrm{pk},\rm i}}{\nu_\mathrm{o,i} F_\mathrm{o,i}}  &=& 
    \begin{cases}
        &\  1 - \widetilde{T}^{-3}_\mathrm{eff1,i} \hspace{2.0cm} \text{for $\widetilde{T}_\mathrm{i} \leq \widetilde{T}_\mathrm{f,i}$}\;,\\
        &\ (1 -\widetilde{T}^{-3}_\mathrm{eff1,i} \lvert_{\widetilde{T}_\mathrm{f,i} })   \; \widetilde{T}^{-3}_\mathrm{eff2,i} \hspace{0.48cm} \text{for $\widetilde{T}_\mathrm{i} \geq \widetilde{T}_\mathrm{f,i}$}\;,\quad\quad \\ 
    \end{cases} \label{inst_spectra} 
    \\ \label{inst_spectra3}
    \frac{\nu_\mathrm{pk,i}}{\nu_\mathrm{o,i}}  &=& \begin{cases}
       &\  \widetilde{T}^{-1}_\mathrm{i} \hspace{1.5cm}\text{for $\widetilde{T}_\mathrm{i} \leq \widetilde{T}_\mathrm{f,i}$}\;,\\
       &\  \widetilde{T}^{-1}_\mathrm{f,i} \; \widetilde{T}^{-1}_\mathrm{eff2,i} \hspace{0.7cm}\text{for $\widetilde{T}_\mathrm{i} \geq \widetilde{T}_\mathrm{f,i}$\;,}
   \end{cases}
\end{eqnarray} 
where the product $\nu_\mathrm{o,i} F_\mathrm{o,i}$ is defined in Appendix \ref{appI} and  $S[x]$ is the normalized Band function defined in equation \ref{Sx}
, $\widetilde{T}_\mathrm{f,i} = R_\mathrm{f,i}/R_\mathrm{o}$, the effective angular timescales are $\widetilde{T}_\mathrm{eff1,i} = (1 - g^2_\mathrm{i}) + g^2_\mathrm{i} \widetilde{T}_\mathrm{i}$ and $\widetilde{T}_\mathrm{eff2,i} = (1 - g^2_\mathrm{i}) + g^2_\mathrm{i} (R_\mathrm{o}/R_\mathrm{f,i}) \widetilde{T}_\mathrm{i}$ (see Appendix \ref{appJ}) such that $\widetilde{T}_\mathrm{eff1,i}\lvert_{\widetilde{T}_\mathrm{f,i}} = (1 - g^2_\mathrm{i}) + g^2_\mathrm{i}R_\mathrm{f,i}/R_\mathrm{o}$. In Appendix \ref{radial_est} we use our model to fit  $\nu_\mathrm{pk} F_\mathrm{\nu_\mathrm{pk}}$ vs. $\nu_\mathrm{pk}$ data for representative GRB samples  from \cite{2023arXiv230800772Y}. Assuming the peak emission being due to RS we find $\Delta R/R_\mathrm{o}$ to be of order unity (see Table I in Appendix \ref{radial_est}).

The right panels of Fig.~\ref{LC_spectra} show the peak of the instantaneous spectrum from each shock steadily rises at early times and decays rapidly at later times. At intermediate times the spectra show a double bump structure which becomes more prominent in the integrated spectra. The low frequency bump is due to FS while the high energy emission is due to RS. While observationally, the low-energy bump is typically interpreted due to be of photospheric origin, our model suggests  a weaker FS as a natural alternative candidate (see \S\,\ref{summary}). 
\begin{figure}
    \centering
    \includegraphics[scale=0.5]{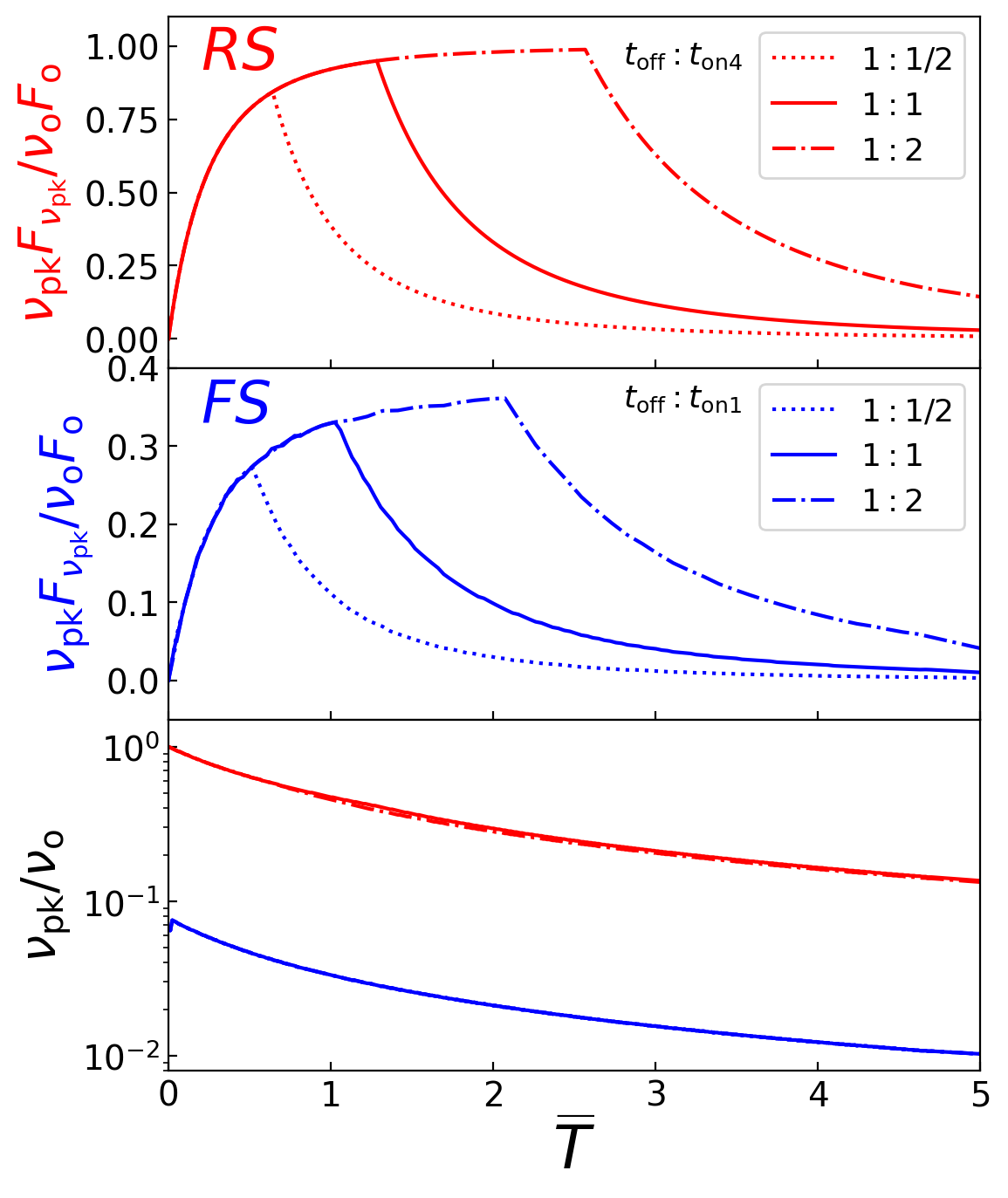}
    \caption{The normalized time evolution of peak luminosity and peak flux. The red and blue lines correspond to the RS and the FS. In the bottom panel the top curves are for RS and the bottom curves are for FS (see text for details).}
     \label{peak_evolution}
\end{figure}

Fig. \ref{peak_evolution} shows that the peak flux of the instantaneous $\nu_{\rm pk,i}F_{\nu, \rm pk,i}$ for the shock front $i$ first rises steadily, then reaches a plateau phase and subsequently decays rapidly, while $\nu_{\rm pk,i}$ shows a monotonic hard-to-soft evolution (see \S \ref{summary}).

\begin{figure}
\includegraphics[scale=0.45]{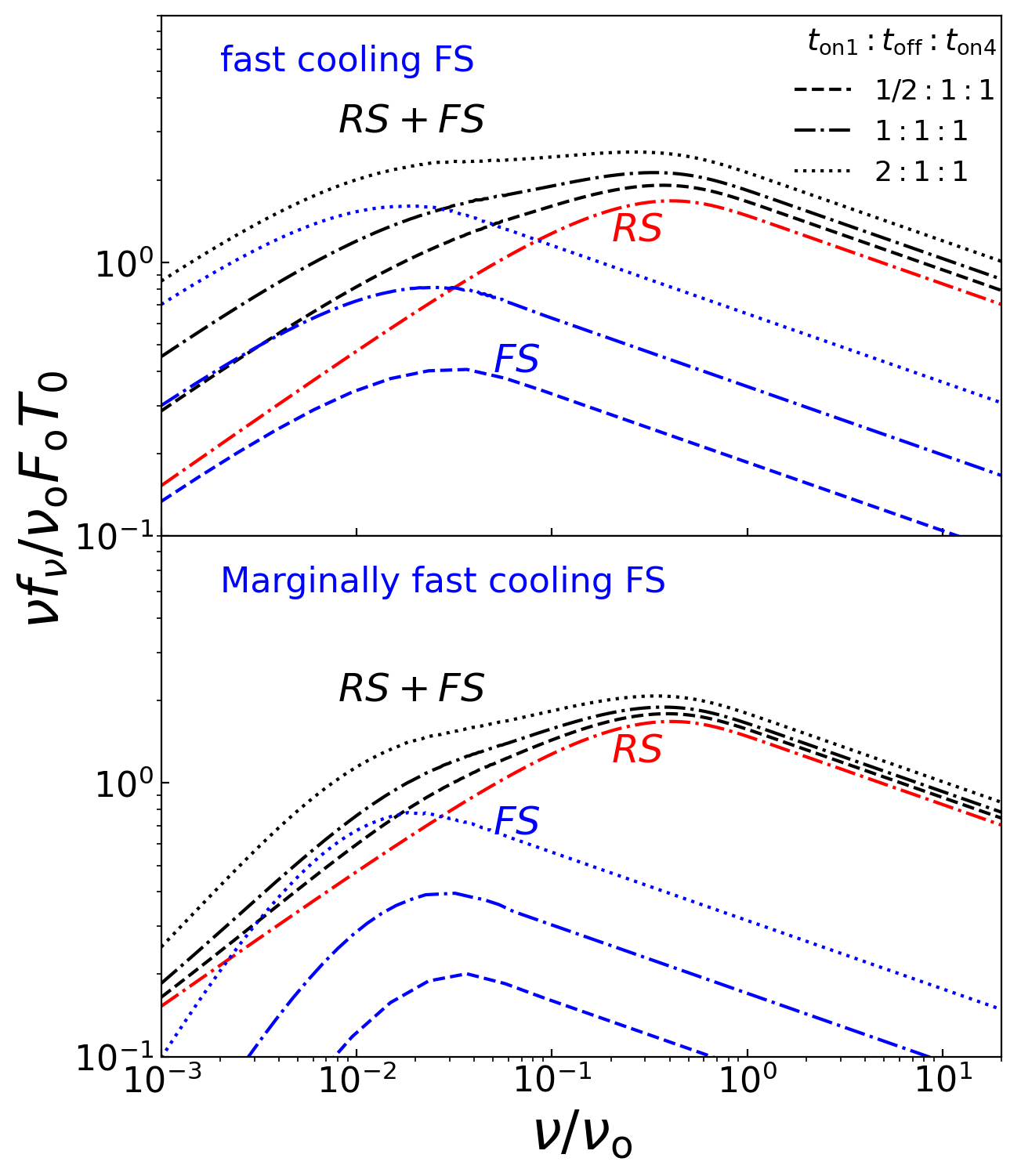}
    \caption{Time integrated spectrum for different combinations of $t_\mathrm{on1}:t_\mathrm{off}:t_\mathrm{on4}$. Here $f_{\nu} = \int dT\,F_{\nu}(T)$ is the fluence per unit frequency. \textbf{\textit{Top}}: the spectrum wherein both RS and FS are in the fast cooling regime. \textbf{\textit{Bottom}}: the spectrum for a fast-cooling RS and a marginally fast cooling FS.}
    \label{integ_spectra}
\end{figure}

Fig. \ref{integ_spectra} shows the time-integrated spectra which are well-modeled (within an error of less than 4\%) as
\begin{eqnarray}
\label{int_spectra1}
(\nu f_{\nu})_\mathrm{i} &=&   \nu_\mathrm{pk,i} f_\mathrm{\nu_\mathrm{pk},i} \;  S[\nu/\nu_\mathrm{pk,i}]\ , \\ \label{int_spectra2}
\frac{\nu_\mathrm{pk,i} f_\mathrm{\nu_\mathrm{pk},i}}{\nu_\mathrm{o,i} F_\mathrm{o,i} T_\mathrm{0,i} \overline{T}_\mathrm{f,i}} &=& 1.32 - 5.62 \times 10^{-3} \log_{10}  \left( \frac{R_\mathrm{f,i}}{R_\mathrm{o}} \right)\ ,
\\ \label{int_spectra3}
\frac{\nu_\mathrm{pk,i}}{\nu_\mathrm{o,i}} &=& \left[0.805 + 0.706 \left(\frac{R_\mathrm{f,i}}{R_\mathrm{o}}\right)\right]^{-1}\ .  
\end{eqnarray}

The top panel shows the spectra for a fast cooling FS. In this regime the overall spectrum is consistent with two spectral break fits viz., a high energy peak and a low energy spectral break. Some studies \cite[e.g.,][]{2021A&A...652A.123T} indeed favor such doubly-broken power-law fits.
The bottom panel shows spectra when the FS is marginally fast cooling. The RS is fast cooling in both panels. It can be seen that the low energy bump becomes more prominent if the FS is longer-lived. In this regime the overall spectra can be well-fit with with a Band-function for the dominant high energy peak and a black-body (BB) like function for the sub-dominant peak. Few studies \cite[e.g.,][]{2011ApJ...727L..33G,2013ApJ...770...32G} favour the (Band-function+BB) fit. Our model accommodates both kind of fits in a natural way. Both the break in the spectral slope at low energies and a sub-dominant bump are due to a weaker FS in different cooling regimes. The higher energy emission (peak) is dominated by RS.

\section{Discussion} \label{summary}

We have presented a self-consistent internal shocks model for the prompt GRB emission, accounting for the dynamics and synchrotron radiation of the forward and reverse shocks (FS, RS), which naturally explains the variability of the lightcurves, the temporal behaviour of the instantaneous spectra and the shape of the time-integrated spectrum. These features are obtained from the
hydrodynamics alone, since the microphysical parameters $(\epsilon_\mathrm{e},\epsilon_\mathrm{B},\xi_\mathrm{e},p)$ are taken to be equal in both shocked regions. Any diversity in those parameters will most likely lead to even more diversity in the observed features. 

Since an internal collision between two shells generically leads to two shocked regions with distinct physical conditions, our results are distinct from one zone synchrotron models for GRB internal shocks \citep{Katz1994,Rees1994,1998MNRAS.296..275D,KM2008,Beniamini2013}. In the single zone model, there is a conflict. On one hand, the prompt emission energetics require a high radiative efficiency $\epsilon_{\rm rad}\,\!\sim\,\!1$. On the other hand, harder observed spectra, $\alpha \approx -1$ (\citealt{Kaneko2006,2011A&A...530A..21N}; \citealt{2012MNRAS.421.1256N}; where $\frac{dN}{dE} \propto E^{\alpha}$), requires the emission to be in the slow or at most in the marginally fast cooling regime. 
This motivated works like \cite[e.g.,][]{2011A&A...526A.110D,2018MNRAS.476.1785B} to consider single zone models involving a marginally fast cooling regime of synchrotron emission with fine tuned parameters to achieve the same. In particular, \citealt{2011A&A...526A.110D}
show that differential IC cooling can 
lead to a low-energy spectral slope that resembles marginally fast-cooling. One example of fine tuning is the requirement to have a balance between heating and cooling of electrons which is difficult to maintain continuously and uniformly. 

Our model has two emission zones, which are linked by the underlying shock hydrodynamics.
In particular, the emission is dominated by the fast cooling RS while the slope at low energies is determined by the FS. Due to the intrinsic weakness of the FS it may naturally be in the slow/marginally fast cooling regime. As it is sub-dominant this does not affect the overall efficiency, which is dominated by the RS. The overall spectrum is the sum
of the emission from the two shocks, and can therefore exhibit a doubly-broken power-law. The lowest-energy power-law index can be $\alpha\approx-2/3$ from a slow cooling FS. The intermediate segment can be $-2\lesssim\alpha\leq-3/2$ (between the peaks of the emission from the two shocks). The highest energy segment is $\alpha=-(p+2)/2$ dominated by the RS. Indeed such spectral models have been successfully fit to prompt GRB data \citep[e.g.,][]{2014ApJ...784...17B,2017ApJ...846..137O,2019A&A...625A..60R}. 

Moreover, for the low-energy bump in the time-integrated spectrum, which is usually
interpreted as a quasi-thermal optically thick ``photospheric''  component
(see \citealt{2017ApJ...846..138G} and the references therein), we find an alternative explanation as an optically thin ``non-thermal'' emission from 
the relatively weaker FS.  The observed spectrum both at the low- and high-energy ends is dominated by the more powerful RS.
It is possible that at least in some cases the low-energy bump has contribution from a photospheric component as well. However, if the contribution from the FS dominates, it implies a sub-dominant photospheric contribution. Observational fits to the time-resolved and time-integrated spectra using equations (\ref{inst_spectra1})-(\ref{inst_spectra3}) and (\ref{int_spectra1})-(\ref{int_spectra3}), respectively, will provide more robust constraints on the ratio of the shock strengths. 

In our model, there is a monotonic hard to soft evolution of the peak photon energy. In Appendix \ref{radial_est}, we show how the $\nu_\mathrm{pk} F_\mathrm{\nu_\mathrm{pk}}$ vs. $\nu_\mathrm{pk}$ data can be exploited 
to infer $\Delta R/R_0$ in our model. However, in some GRBs $\nu_\mathrm{pk}$ follows $\nu_\mathrm{pk}F_\mathrm{\nu_\mathrm{pk}}$, referred to as intensity tracking \citep{1983Natur.306..451G,1996Natur.381...49L,2012ApJ...756..112L}. \citet{Beniamini-Granot-16} show that intensity tracking can be reproduced when the GRB energy dissipation is driven by magnetic reconnection. Thus, intensity-tracking may point to a different underlying physics. 

One may potentially distinguish between
a true photospheric emission and a non-thermal photospheric-like emission from the FS. The true photospheric emission arises much closer to the source, leading to an earlier onset time -- a precursor emission to each pulse, which may be detectable in particularly bright pulses. It will also have a sharper spectral peak and a harder low-energy photon index.

Finally, we point out a few limitations of our current study. 
We have assumed a moderate
proper speed contrast $(a_\mathrm{u} = 2)$, which roughly reproduces the observed ratio of the peak photon energies, $E_\mathrm{pk,RS}/E_\mathrm{pk,FS}\sim10-10^{1.5}$. However,  this ratio scales as the square of the shock strengths, $E_\mathrm{pk,RS}/E_\mathrm{pk,FS} \propto (\Gamma_{34} -1)^2/(\Gamma_{21 - 1})^2$, and can become very large for $a_\mathrm{u} \gg 1$, which is not observed.  However, this may be mitigated by accounting for the affects of the spherical geometry on the shock dynamics, which are expected to reduce the ratio of shock strengths and thereby also $E_\mathrm{pk,RS}/E_\mathrm{pk,FS}$ with $R/R_\mathrm{o}$ and time. The thin shell instantaneous emission region approximation may also break for a marginally fast cooling FS. Shock propagation in a spherical geometry, larger proper speed contrasts and a finite instantaneous emission region will be pursued in a future study.

\section*{Acknowledgements}

This research was funded in part by the ISF-NSFC joint research program under grant no. 3296/19 (S.M.R., J.G.) and by the United States-Israel Binational Science Foundation (BSF) under grant no. 2020747 (P.B.). We thank Frederic Daigne and Robert Mochkovitch for useful discussions and the referee for useful comments.
\section*{Data Availability}

No new data were generated during the analysis of this project. 



\bibliographystyle{mnras}
\bibliography{References} 




\appendix

\section*{Online-only supplementary material}

The Appendices are available as online only supplementary material.

\onecolumn

\appendix

\section{Solving for the proper speed of the shocked fluid in the ultra-relativistic limit} \label{appA}

From \cite{rahaman2023internal} (hereafter Paper I) for the collision of two cold shells , the \cite{1976PhFl...19.1130B} conditions governing shock hydrodynamics gives,
\begin{equation}
    (\Gamma^2_{21} - 1) = f (\Gamma^2_{34} - 1)
\end{equation}

Using the ultra-relativistic approximation (see Eq. 3 of \citealt{1995ApJ...455L.143S}) we have 
\begin{eqnarray}
     & \Gamma_{21} \approx \frac{1}{2} \left( \frac{\Gamma_{2}}{\Gamma_{1}} + \frac{\Gamma_{1}}{\Gamma_{2}} \right) = \frac{1}{2} \left( \frac{\Gamma }{\Gamma_{1}} + \frac{\Gamma_{1}}{\Gamma } \right)  \\
     & \Gamma_{34} \approx \frac{1}{2} \left( \frac{\Gamma_{3}}{\Gamma_{4}} + \frac{\Gamma_{4}}{\Gamma_{3}} \right)  = \frac{1}{2} \left( \frac{\Gamma }{\Gamma_{4}} + \frac{\Gamma_{4}}{\Gamma } \right)  
\end{eqnarray}

Using the ultra-relativistic approximation $\Gamma_4/\Gamma_1 \approx u_4/u_1 = a_\mathrm{u}$ and plugging equations (A2)-(A3) in equation (A1)
we get,
\begin{equation}
    \Gamma = \sqrt{\frac{\sqrt{f} a^2_\mathrm{u} + a_\mathrm{u}}{a_\mathrm{u} + \sqrt{f}}} \Gamma_1 \approx \frac{\sqrt{2} a_\mathrm{u} \Gamma_1}{\sqrt{1 + a^2_\mathrm{u}}} \quad \quad \text{for constant source power $L$ during shell ejection}
\end{equation}

\section{Estimation of thermal efficiencies}\label{appD}

The thermal efficiency (defined as the ratio of the internal energy dissipated in region $j$ to the total kinetic energy available in both shells before collision) in region R2 and R3 is given by (see Eq. 38a and 38b of Paper I)
\begin{equation}
    \epsilon_\mathrm{th,j} = \alpha_\mathrm{j}  \frac{E_\mathrm{int,j,max}}{E_\mathrm{k,1,0}+E_\mathrm{k,4,0}} = \frac{E_\mathrm{int,j}}{E_\mathrm{k,1,0}+E_\mathrm{k,4,0}} \hspace{2cm} \text{For the combination ij = (12,34) }
\end{equation}
where $\alpha_\mathrm{j}$ is unity if the shock finishes crossing region $i$ (see Appendix D of Paper I) such that
\begin{eqnarray}
    &\ E_\mathrm{int2,max} = M_{1} c^2 \Gamma  \left[1 + \beta^2 \left(\frac{\Gamma_{21} + 1}{3 \Gamma_{21}}\right) \right] (\Gamma_{21} - 1) = \left( \frac{L t_\mathrm{on1}}{\Gamma_1 - 1} \right) \Gamma \left[1 + \beta^2 \left(\frac{\Gamma_{21} + 1}{3 \Gamma_{21}}\right) \right] (\Gamma_{21} - 1)                          \\
    &\  E_\mathrm{int3,max} =  M_{4} c^2 \Gamma \left[1 + \beta^2 \left( \frac{\Gamma_{34} + 1}{3 \Gamma_{34}}\right) \right] (\Gamma_{34} - 1)   = \left( \frac{L t_\mathrm{on4}}{\Gamma_4 - 1} \right) \Gamma \left[1 + \beta^2 \left( \frac{\Gamma_{34} + 1}{3 \Gamma_{34}}\right) \right] (\Gamma_{34} - 1)   
\end{eqnarray}
assuming constant source power $L$ during shell ejection. 

One can estimate an alternative quantity $\epsilon_\mathrm{i}$ (defined as the internal energy dissipated in shell $i$ to the available kinetic energy in that shell before collision) expressed as
\begin{equation}
    \epsilon_\mathrm{i} = \frac{E_\mathrm{int,j}}{E_\mathrm{k,i,0}} = \frac{\Gamma }{\Gamma_\mathrm{i} - 1}  \left[ 1 + \beta^2 (\hat{\gamma}_\mathrm{j}-1) \right] (\Gamma_\mathrm{ij} - 1) \approx \frac{\hat{\gamma}_\mathrm{j}}{2} \left( \frac{\Gamma }{\Gamma_\mathrm{i}} - 1\right)^2  \hspace{2cm} \text{For the combination ij = (12,34)}
\end{equation}
where the adiabatic constant $\hat{\gamma}_\mathrm{j} = \frac{4 \Gamma_\mathrm{ij} + 1}{3 \Gamma_\mathrm{ij}}$ (\citealt{2003ApJ...591.1075K}), and we have used the ultra-relativistic approximations $\beta \sim 1$, $\Gamma_\mathrm{i} - 1 \approx \Gamma_\mathrm{i}$, $\Gamma_\mathrm{ij} \approx  (\Gamma^2 + \Gamma^2_\mathrm{i})/(2 \Gamma \Gamma_\mathrm{i})$. This definition has the unique advantage that it depends only on the shock hydrodynamics. The combined thermal efficiency from hydrodynamics is given as
\begin{equation}
    \epsilon_\mathrm{th,hydro} = \frac{t_\mathrm{on1}}{t_\mathrm{on1} + t_\mathrm{on4}} \epsilon_{1} + \frac{t_\mathrm{on4}}{t_\mathrm{on1} + t_\mathrm{on4}} \epsilon_{4} = f_\mathrm{on1} \epsilon_1 + f_\mathrm{on4} \epsilon_{4} , \label{therm_hydro} 
\end{equation}

The thermal efficiency for the plastic collision of two-point masses $(M_{1},M_{4})$ of LF $(\Gamma_1,\Gamma_4)$ and initial kinetic energies $(E_{\rm k,1,0} , E_{\rm k,4,0})$ is given by
\begin{equation}
 \epsilon_\mathrm{th,ballistic} = \frac{E_\mathrm{int}}{E_{\rm k,1,0} + E_{\rm k,4,0}} 
 = 1 - \frac{(\Gamma_\mathrm{ballistic} -1)\left( 1 + \frac{M_{4}}{M_{1}} \right)}{(\Gamma_1 - 1) + (\Gamma_4 -1) \frac{M_{4}}{M_{1}} }\ , \label{therm_ball}
\end{equation}
where
\begin{align}
    &\ \frac{M_4}{M_1} = \frac{\frac{E_\mathrm{k,4,0}}{\Gamma_\mathrm{4} - 1}}{\frac{E_{k,1,0}}{\Gamma_{1} - 1}} = \frac{t_\mathrm{on4}}{t_\mathrm{on1}} \left( \frac{\Gamma_1 - 1}{\Gamma_4 -1}\right) \approx \frac{t_\mathrm{on4}}{t_\mathrm{on1}} \frac{\Gamma_1}{\Gamma_4}  \quad \quad \text{For constant source power $L$ and $(u_1,u_4) \gg 1$}  \\ 
    &\ \Gamma_\mathrm{ballistic} = \frac{\Gamma_{1}  + \Gamma_{4} \frac{M_{4}}{M_{1}} }{\sqrt{1 + \frac{M^2_{4}}{M^2_{1}} + 2 \Gamma_{41} \frac{M_{4}}{M_{1}}} } \\
    &\ \Gamma_{41} = \Gamma_1 \Gamma_4 (1 - \beta_1 \beta_2) 
\end{align}

We define a dimensionless ratio $\eta = t_\mathrm{on1}/t_\mathrm{on4}$ as the ratio of the ejection timescales for shells S1 and S4. 

\begin{figure}
    \begin{minipage}[c]{0.3\linewidth}
       \hspace{0.25cm} \includegraphics[scale=0.54]{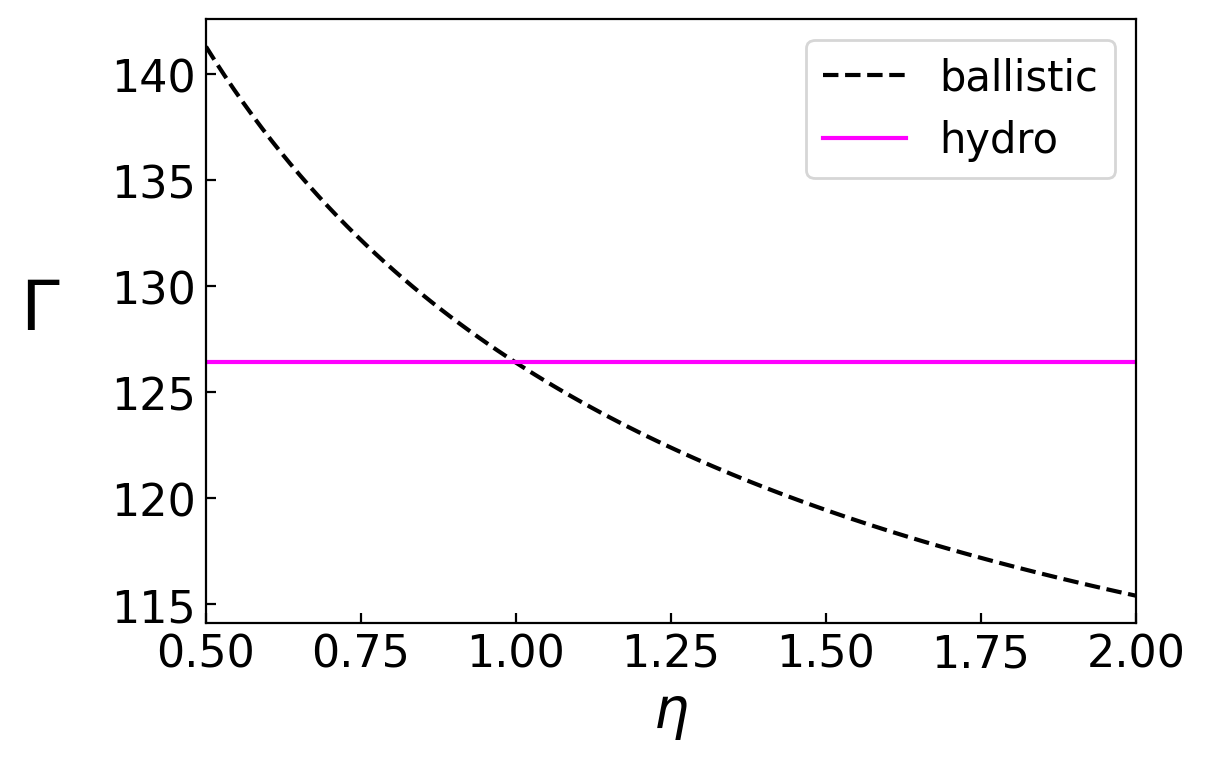}\\
        \hspace{-6cm} \includegraphics[scale = 0.47]{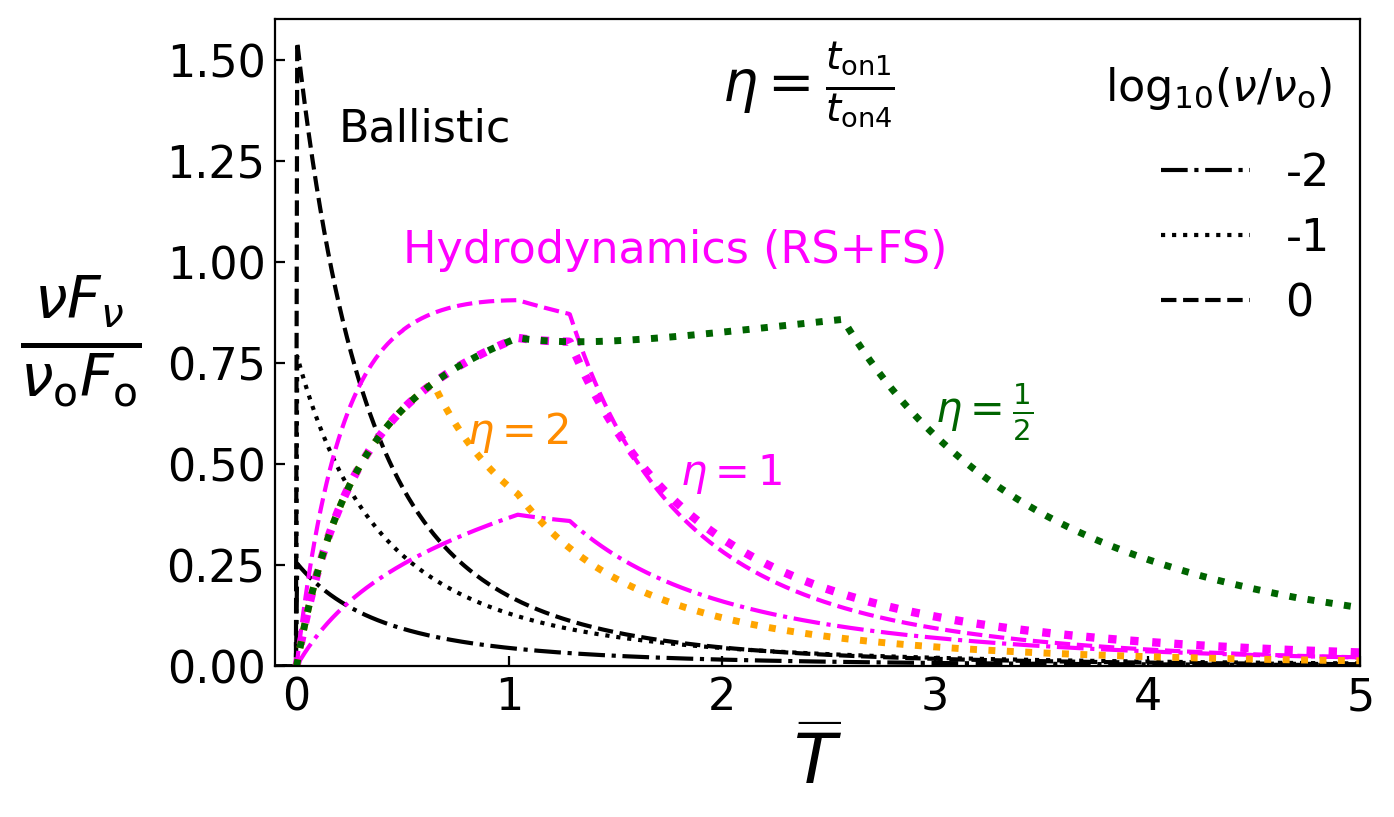}
    \end{minipage}\hfill 
     \begin{minipage}[c]{0.53\linewidth}
     \includegraphics[scale=0.58]{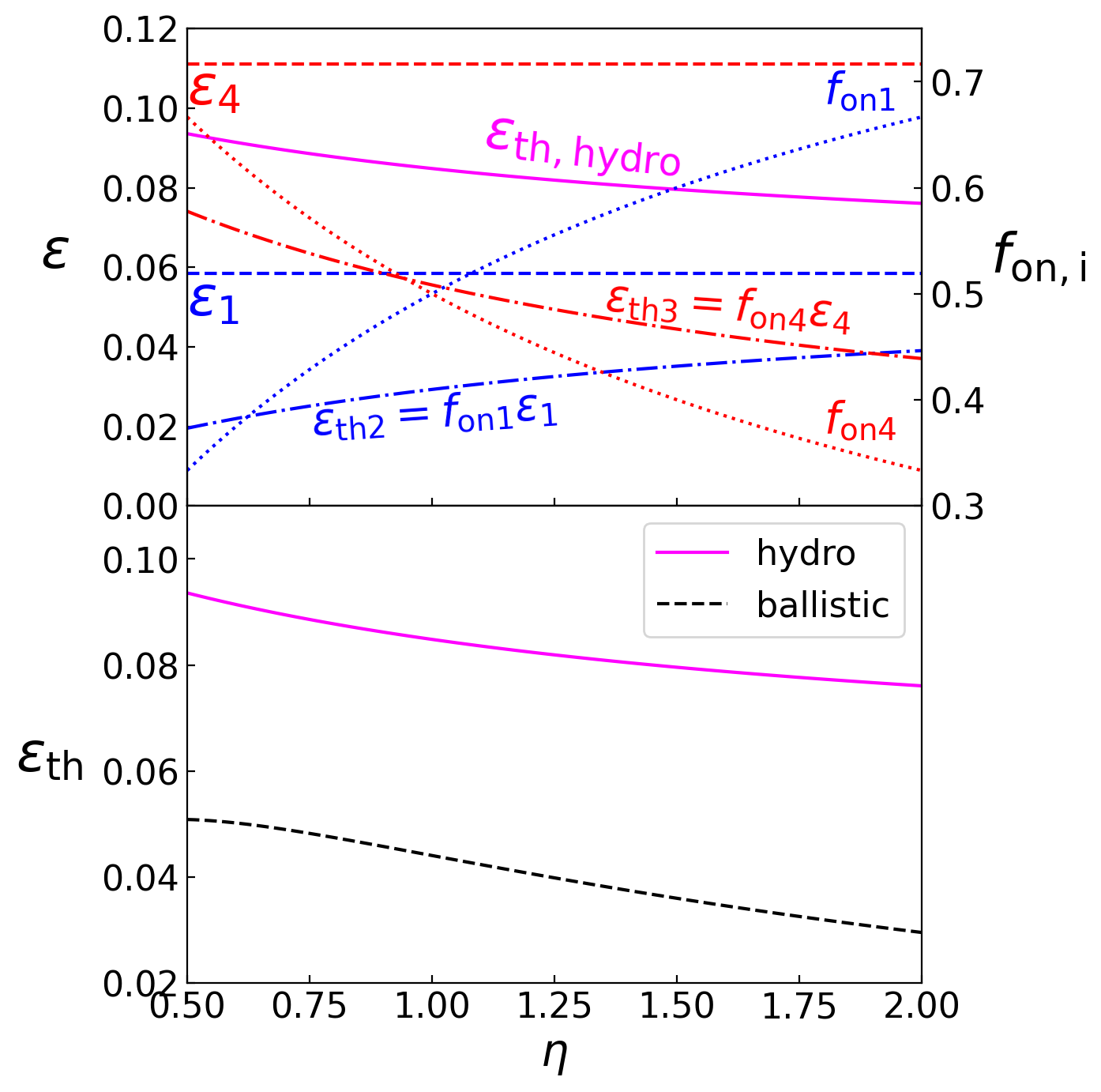}
    \end{minipage}
    \caption{Comparison of the ballistic and the hydrodynamic collision of two  shells moving with proper speeds $(u_1,u_4) = (100,200)$. \textbf{Left:}The upper panel and the lower panel compares the LF and the lightcurves. The LF of the shocked material is same for both cases at $\eta = 1$ (corresponding to collision of equal energy shells). The lightcurves are shown at three normalized frequencies for the ballistic case (in black) and hydrodynamic case (shown in magenta) for $\eta =1$. The orange and green dotted lines correspond to hydrodynamic lightcurves at a fixed normalized frequency for two different $\eta = (2,1/2)$. The lightcurves satisfies follows the normalization $(\int d\nu f_{\nu}  )_\mathrm{ballistic}/(\int d \nu f_{\nu}  )_\mathrm{hydro} = \epsilon_\mathrm{th,ballistic}/\epsilon_\mathrm{th,hydro}$ where $f_{\nu}$ is the fluence per unit frequency. \textbf{Right:} shows the thermal efficiencies $(\epsilon_\mathrm{th,ballistic}, \epsilon_\mathrm{th,hydro})$ as a function of $\eta$ (see text for detailed explanation). }
    \label{hydro_ball_compare}
\end{figure}

The ballistic approach assumes a plastic collision for infinitely thin shells.  The infinitely thin shell approximation is akin to considering collision of effectively two point masses. For a fixed proper speed of the colliding shells $(u_1,u_4)$ quantities in the ballistic approach depend only on the ratio $M_4/M_1$. Fig. \ref{hydro_ball_compare} gives compares the features from the ballistic and the hydrodynamic approaches for a constant source power $L$ of the compact source. The upper panel on the left shows that while  the LF of the shocked material in the hydrodynamic case is constant as the proper density contrast $f$ is constant. While the LF of the merged shell for ballistic case changes as a function of $\eta$ (which corresponds to a changing $M_4/M_1$ ratio). The LF of the shocked material matches exactly for both cases for collision of two equal energy shells. The lower panel on the left compares the lightcurves for ballistic approach (shown in black) and the hydrodynamic approach (shown in magenta). The peak comoving frequency $\nu_\mathrm{o}$ for both approaches have been taken to be  the same (which we take to be $\nu_\mathrm{o,RS}$). The ratio of the frequency integrated fluence for both approaches has been taken to equal to the ratio of the thermal efficiencies defined in equations \ref{therm_hydro} and \ref{therm_ball} respectively. The expression for the light curve for the ballistic case (emission from a single collision radii $R_\mathrm{o}$) is taken from \citet{2009MNRAS.399.1328G}. In the hydrodynamic case for a fixed observed frequency, the shape of the light curve changes if the ratio $\eta$ changes shown in dotted orange at a lower value of $\eta$ and in dotted green for a larger value of $\eta$. Thus, the hydrodynamic approach not only accounts for different pulse shapes at different frequencies but accounts for the change of pulse shape at a fixed frequency. It has the following observational implication: means that the change of pulse morphology across frequencies (for a single burst) and change of pulse at a fixed frequency (from burst to burst) can be accounted by the hydrodynamic approach. The right panel shows that the even when thermal efficiencies are compared even for our moderate proper speed contrast of $a_\mathrm{u} = 2$, the thermal efficiency for the hydrodynamic approach is almost twice as large as the ballistic approach. This means that the traditional criticism that thermal efficiency for internal shock stays small can be circumvented if one takes into account hydrodynamic approach. We refer the reader to Paper I for an in-depth treatment of hydrodynamic thermal efficiency.

\section{Ratio of the peak frequencies at $(R_\mathrm{\lowercase{o}},\theta=0)$ in two shocked regions}\label{appE}

\begin{table*}
    \centering
    \caption{Symbols $\&$ definitions to describe microphysical parameters of the synchrotron model}
    \begin{tabular}{cc} \hline
    Symbol & Definitions\\ \hline 
      $\epsilon_\mathrm{B}$  & The fraction of the internal energy density that goes into magnetic field energy density \\
      $\epsilon_\mathrm{e}$ & The fraction of the internal energy density that goes into non-thermal energy density \\ 
      $\xi$   & Fraction of the electrons accelerated to non-thermal distribution \\ 
      $p$ & power-law index of the accelerated electrons \\ \hline  
    \end{tabular}  \label{micro_param}
\end{table*}

In the comoving frame the minimum Lorentz factor of the non-thermal electrons for $p>2$ is
\begin{equation}
    \gamma_{\rm m} = G(p)   \,\frac{m_\mathrm{p}}{m_\mathrm{e}}\,\frac{\epsilon_\mathrm{e}}{\xi_\mathrm{e}}\,  (\Gamma_\mathrm{ij} - 1) \ , 
\end{equation}
where $\Gamma_\mathrm{ij}$ is the relative upstream to downstream Lorentz factor across the shock, $G(p) = (p-2)/(p-1)$ and $p$ is the power-law slope of the non-thermal electrons. 
The comoving magnetic field strength can be estimated as 
\begin{equation}
    B' = \sqrt{\epsilon_\mathrm{B} 8 \pi e'_\mathrm{int}} = \sqrt{\epsilon_\mathrm{B} \frac{ 8 \pi P}{\hat{\gamma} - 1}} = \sqrt{ \epsilon_\mathrm{B} 24 \pi P} \sqrt{\frac{\Gamma_\mathrm{ij}}{\Gamma_\mathrm{ij} + 1}}
\end{equation}
where $P$ is the gas pressure in the shocked region and we assume the adiabatic constant of the heated gas to be $\hat{\gamma} = (4 \Gamma_\mathrm{ij} +1)/3\Gamma_\mathrm{ij}$ following \cite{2003ApJ...591.1075K}.  

The peak frequency in the comoving frame is given by
\begin{equation}
    \nu'_\mathrm{m}  \propto  B'\gamma^2_\mathrm{m}  \propto  G^2(p) \epsilon^2_\mathrm{e} \epsilon_\mathrm{B}^{1/2}\xi^{-2}_\mathrm{e} (\Gamma_\mathrm{r} - 1)^2 \sqrt{\frac{P}{\hat{\gamma} - 1}} \propto  
    \sqrt{P} \; G^2(p) 
    \epsilon^2_\mathrm{e} \epsilon_\mathrm{B}^{1/2} \xi^{-2}_\mathrm{e} (\Gamma_\mathrm{ij} - 1)^2 \sqrt{\frac{\Gamma_\mathrm{ij}}{\Gamma_\mathrm{ij} + 1}}\ .
\end{equation}

In the shock hydrodynamic condition we assume that the pressure in both shock regions are equal. Additionally we assume, the microphysical parameters $(\epsilon_\mathrm{e}, \epsilon_\mathrm{B},\xi_\mathrm{e},p)$ to be the same in both shocked regions.  Thus, the ratio of the peak frequencies in the two shocked regions is given as 
\begin{equation}
    \frac{\nu'_\mathrm{m,RS}}{\nu'_\mathrm{m,FS}} = \frac{\nu'_\mathrm{o,RS}}{\nu'_\mathrm{o,FS}} =\left( \frac{\Gamma_{34} - 1}{\Gamma_{21}-1} \right)^2 \left[ \frac{\Gamma_{34}}{\Gamma_{21}} \left( \frac{\Gamma_{21} + 1}{\Gamma_{34} + 1}\right) \right]^{1/2} =  \left( \frac{\Gamma_{34} + 1}{\Gamma_{21} + 1} \right)^{-1/2} \left( \frac{\Gamma_{34}}{\Gamma_{21}} \right)^{1/2} \left( \frac{\Gamma_{34} - 1}{\Gamma_{21} - 1} \right)^2\ . \label{Ratio_freq}
\end{equation}

For constant source power $L$ and in the ultra-relativistic limit $(u_1,u_4) \gg 1$ we have
\begin{equation}
    \nu'_\mathrm{o,RS} = \nu'_\mathrm{m,RS} = \gamma^2_\mathrm{m} \frac{q_\mathrm{e} B'}{2 \pi m_\mathrm{e} c} = \frac{q_\mathrm{e}}{\sqrt{2} \pi m_\mathrm{e} c^{5/2}} \left( \frac{m_\mathrm{p}}{m_\mathrm{e}}\right)^2 \epsilon^{1/2}_{\rm B} \epsilon^2_{\rm e} \xi^{-2}_{\rm e} t^{-1}_\mathrm{off} L^{1/2} (\Gamma_{34} - 1)^{5/2} (u_1 a_\mathrm{u})^{-3} (a^2_\mathrm{u} - 1)
\end{equation}

We have used $\nu_\mathrm{o} = \nu_\mathrm{o,RS} = 2 \Gamma \nu'_\mathrm{o,RS}/(1+z)$ for frequency normalization for all figures (for display purposes) in the main text. Eq. (G18) in Appendix G gives typical value of $\nu_\mathrm{o}$ in terms of fiducial microphysical  and central engine parameters.

\section{Ratio of the peak luminosities at $(R_\mathrm{\lowercase{o}},\theta=0)$ in the two shocked regions}\label{appF}

The comoving bolometric luminosity from the fast synchrotron model is given as
\begin{equation}
    L'_\mathrm{bol,\gamma} = W(p) \sqrt{\nu'_\mathrm{m} \nu'_\mathrm{c}} L'_\mathrm{\nu',max} 
    = W(p) \nu'_\mathrm{m} L_{\nu'_\mathrm{m}} 
    = W(p) \left[ \nu'_\mathrm{o} \left( \frac{R}{R_\mathrm{o}}\right)^d \right] \left[ L'_\mathrm{o} \left( \frac{R}{R_\mathrm{o}}\right)^a \right] 
    = W(p) v'_\mathrm{o} L'_\mathrm{o} \left( \frac{R}{R_\mathrm{o}} \right)^{a+d}\ ,  
\end{equation}
where $W(p) = 2(p-1)/(p-2)$ is a factor of order unity, while  $\nu'_\mathrm{o} = \nu'_\mathrm{m}(R=R_\mathrm{o})$ and $L'_\mathrm{o} = L'_{\nu'_\mathrm{m}} (R = R_\mathrm{o})$. For the internal shock model we have $a+d=0$ which gives 
\begin{eqnarray}
    L'_\mathrm{bol,\gamma} = W(p) \nu'_\mathrm{o} L'_\mathrm{o}\ .
\end{eqnarray}
Assuming the same $W(p)$ in both shocked regions, the ratio of the comoving bolometric luminosities in  the reverse and the forward shocked region is given as
\begin{equation}
    \frac{L'_\mathrm{bol,RS}}{L_\mathrm{bol,FS}} = \left( \frac{\nu'_\mathrm{o,RS}}{\nu'_\mathrm{o,FS}} \right) \left( \frac{L'_\mathrm{o,RS}}{L'_\mathrm{o,FS}} \right) \label{Ratio1_Lbol}
\end{equation}

Next, we can represent the comoving bolometric luminosity from shock hydrodynamics as
\begin{equation}\label{bol_shock}
    L'_\mathrm{bol,\gamma} = \epsilon_\mathrm{rad} \left( \epsilon_\mathrm{e}  \frac{dE'_\mathrm{int}}{dt'}\right) =  \epsilon_\mathrm{\gamma} \frac{dE'_\mathrm{int}}{dt'} = \epsilon_\mathrm{\gamma} (\Gamma_\mathrm{ij} - 1) \frac{dM}{dt'} c^2 \ ,
\end{equation}
where $\epsilon_\mathrm{rad} = \epsilon_{e} \epsilon_{\gamma} $.
The rate of change of (rest) mass in the reverse shocked region is given by
\begin{equation}\label{M3_rate}
    \frac{dM_{3}}{dt'} = \Gamma_{34} \rho'_4 \left[ \beta_\mathrm{RS,CD} + \beta_{34}\right] c  \; 4 \pi R^2 = \Gamma_{34} \; \rho'_{4,0}  \left( \frac{R_\mathrm{o}}{R} \right)^2 \left[ \beta_\mathrm{RS,CD} + \beta_{34} \right] c \; 4 \pi R^2 = \Gamma_{34} \rho'_{4,0} \left[ \beta_\mathrm{RS,CD} + \beta_{34} \right] c \; 4 \pi R^2_\mathrm{o}\ .  
\end{equation} 
Similarly, the rate of change of mass in the forward shocked region is given as
\begin{equation}\label{M2_rate}
    \frac{dM_{2}}{dt'} = \Gamma_{21} \rho'_{1,0} \left[ \beta_\mathrm{FS,CD} + \beta_{21} \right] c\;  4 \pi R^2_\mathrm{o}\ .  
\end{equation}
Using equations (\ref{M3_rate}) and (\ref{M2_rate}) in equation (\ref{bol_shock}) we obtain,
\begin{eqnarray}\label{Lbol_RS}
    &\ L'_\mathrm{bol,RS} = \epsilon_\mathrm{\gamma,RS} (
    \Gamma_{34} - 1) \; \Gamma_{34} \rho'_{4,0} \left[ \beta_\mathrm{RS,CD} + \beta_{34} \right] c^3 \; 4 \pi R^2_\mathrm{o}\ = \epsilon_\mathrm{e} \epsilon_\mathrm{rad,RS} (
    \Gamma_{34} - 1) \; \Gamma_{34} \rho'_{4,0} \left[ \beta_\mathrm{RS,CD} + \beta_{34} \right] c^3 \; 4 \pi R^2_\mathrm{o}\ .  \\  \label{Lbol_FS}  
    &\ L'_\mathrm{bol,FS} = \epsilon_\mathrm{\gamma,FS}  (
    \Gamma_{21} - 1) \; \Gamma_{21} \rho'_{1,0} \left[ \beta_\mathrm{FS,CD} + \beta_{21} \right] c^3 \; 4 \pi R^2_\mathrm{o}\ = \epsilon_\mathrm{e} \; \epsilon_\mathrm{rad,FS} (
    \Gamma_{21} - 1) \; \Gamma_{21} \rho'_{1,0} \left[ \beta_\mathrm{FS,CD} + \beta_{21} \right] c^3 \; 4 \pi R^2_\mathrm{o}\ . 
\end{eqnarray}

Assuming the same $\epsilon_\mathrm{e}$ across both the shocked regions, using equations (\ref{Lbol_RS}) and (\ref{Lbol_FS}) the ratio of the bolometric luminosity from both the shocked regions is given by
\begin{equation}
\begin{split}
    \frac{L'_\mathrm{bol,RS}}{L'_\mathrm{bol,FS}} &\ = \left( \frac{\epsilon_\mathrm{\gamma,RS}}{\epsilon_\mathrm{\gamma,FS}} \right)\;  \left( \frac{\rho'_{4,0}}{\rho'_{1,0}}\right) \left( \frac{\Gamma_{34} - 1}{\Gamma_{21} - 1} \right) \left( \frac{\Gamma_{34}}{\Gamma_{21}} \right) \left( \frac{\beta_\mathrm{RS,CD} + \beta_{34}}{\beta_\mathrm{FS,CD} + \beta_{21}} \right) = \left( \frac{\epsilon_\mathrm{\gamma,RS}}{\epsilon_\mathrm{\gamma,FS}} \right)\; \left[ \frac{\Gamma^2_{21}-1}{\Gamma^2_{34} - 1} \right] \left( \frac{\Gamma_{34} - 1}{\Gamma_{21} - 1} \right) \left( \frac{\Gamma_{34}}{\Gamma_{21}} \right) \left( \frac{\beta_\mathrm{RS,CD} + \beta_{34}}{\beta_\mathrm{FS,CD} + \beta_{21}} \right) \\
    &\ = \left( \frac{\epsilon_\mathrm{\gamma,RS}}{\epsilon_\mathrm{\gamma,FS}} \right)\; \left( \frac{\Gamma_{34} + 1}{\Gamma_{21} + 1} \right)^{-1} \left( \frac{\Gamma_{34}}{\Gamma_{21}} \right) \left( \frac{\beta_\mathrm{RS,CD} + \beta_{34}}{\beta_\mathrm{FS,CD} + \beta_{21}} \right) \label{Ratio2_Lbol}
\end{split}
\end{equation}
where we have used the relation $\rho'_{4,0}/\rho'_{1,0} = f = (\Gamma^2_{21} - 1)/(\Gamma^2_{34} - 1)$.

 Equating equations (\ref{Ratio1_Lbol}) and (\ref{Ratio2_Lbol}) and using equation (\ref{Ratio_freq}) we get the ratio of the peak comoving luminosities,
\begin{equation}
    \frac{L'_\mathrm{o,RS}}{L'_\mathrm{o,FS}} = \left( \frac{\nu'_\mathrm{o,RS}}{\nu'_\mathrm{o,FS}} \right)^{-1} \left( \frac{\epsilon_\mathrm{rad,RS}}{\epsilon_\mathrm{rad,FS}} \right)\;\left[ \left( \frac{\Gamma_{34} + 1}{\Gamma_{21} + 1} \right)^{-1} \left( \frac{\Gamma_{34}}{\Gamma_{21}} \right) \left( \frac{\beta_\mathrm{RS,CD} + \beta_{34}}{\beta_\mathrm{FS,CD} + \beta_{21}} \right) \right]  = \left( \frac{\epsilon_\mathrm{rad,RS}}{\epsilon_\mathrm{rad,FS}} \right)\; \left( \frac{\Gamma_{34} + 1}{\Gamma_{21} + 1} \right)^{-1/2}  \left( \frac{\Gamma_{34}}{\Gamma_{21}} \right)^{1/2} \left( \frac{\Gamma_{34} - 1}{\Gamma_{21} - 1} \right)^{-2} \left( \frac{\beta_\mathrm{RS,CD} + \beta_{34}}{\beta_\mathrm{FS,CD} + \beta_{21}} \right)   
\end{equation}

It can be shown that $\beta_\mathrm{RS,CD}+ \beta_{34} = \frac{4}{3} \beta_{34}$ and $\beta_\mathrm{FS,CD}+ \beta_{21} = \frac{4}{3} \beta_{21}$. This further simplifies the previous expression to 
\begin{equation}
     \frac{L'_\mathrm{o,RS}}{L'_\mathrm{o,FS}} =  \left( \frac{\epsilon_\mathrm{rad,RS}}{\epsilon_\mathrm{rad,FS}} \right)\; \left( \frac{\Gamma_{34} + 1}{\Gamma_{21} + 1} \right)^{-1/2}  \left( \frac{\Gamma_{34}}{\Gamma_{21}} \right)^{1/2} \left( \frac{\Gamma_{34} - 1}{\Gamma_{21} - 1} \right)^{-2} \left( \frac{\beta_{34}}{\beta_{21}} \right)   
\end{equation}

\section{EATs of the two shocked regions} \label{appG}

\begin{table}
    \centering
     \caption{Symbols and definitions for physical quantities for describing EATS. The time in small alphabet $t$ represents lab frame time and the time $T$ in the capital alphabet represents the observer frame time.  }
    \begin{tabular}{c|c} \hline 
        Symbol      &  Definition \\ \hline
        $t_\mathrm{ej,f1}$ &  the time when the front edge of shell S1 is ejected \\
         $t_\mathrm{ej,b1}$ &  the time when the back edge  of shell S1 is ejected \\
          $t_\mathrm{ej,f4}$ &  the time when the front of shell S4 is ejected \\
        $ t_\mathrm{off}$ & time delay between the ejection of back edge of S1 and the ejection of front edge of S4 \\ 
        $t_\mathrm{o}$ & The time when shells S1 and S4 collide \\ 
        $R_\mathrm{o}$ & The distance from the central engine where the shells S1 and S4 collide \\
        $t^\mathrm{eff}_\mathrm{ej,FS}$ & The effective lab frame time if shell S1 were ejected at $u_\mathrm{FS}$ instead of $u_1$ \\
        $T^\mathrm{eff}_\mathrm{ej,FS}$ & The (hypothetical) observer frame time when the photons emitted at $t^\mathrm{eff}_\mathrm{ej,FS}$ reach the observer
        \\
        $t^\mathrm{eff}_\mathrm{ej,RS}$ & The effective lab frame time if shell S4 were ejected at $u_\mathrm{RS}$ instead of $u_1$ \\
        $T^\mathrm{eff}_\mathrm{ej,RS}$ & The (hypothetical) observer frame time when the photons emitted at $t^\mathrm{eff}_\mathrm{ej,RS}$ reach the observer \\ \hline 
    \end{tabular}\label{table_EATs}
\end{table}

The notation for the ejection timescales are summarized in Table \ref{table_EATs}. 

The lab frame time $t_\mathrm{o}$ where shells S1 and S4 collide are 
\begin{equation}
    t_\mathrm{o} = t_\mathrm{ej,f4} + \frac{v_1  t_\mathrm{off}}{v_{4} - v_{1}} = t_\mathrm{ej,f4} + \frac{\beta_1  t_\mathrm{off}}{\beta_4 - \beta_1} 
\end{equation}

The distance $R_\mathrm{o}$ from the central engine where both shells (S1,S4) collide are given as
\begin{equation}
    R_\mathrm{o} = v_{4} \left(  \frac{\beta_1  t_\mathrm{off}}{\beta_4 - \beta_1} \right) = \frac{\beta_1 \beta_4 c  t_\mathrm{off}}{\beta_4 - \beta_1}
\end{equation}

We define effective ejection timescales of the shells as 
\begin{eqnarray}
    &\ t^\mathrm{eff}_\mathrm{ej,FS} = t_\mathrm{o} - \frac{R_\mathrm{o}}{\beta_\mathrm{FS} c} = t_\mathrm{ej,f4} + \frac{\beta_1  t_\mathrm{off}}{\beta_4 - \beta_1} \left( 1 - \frac{\beta_4}{\beta_\mathrm{FS}} \right) \\ 
    &\ t^\mathrm{eff}_\mathrm{ej,RS} = t_\mathrm{o} - \frac{R_\mathrm{o}}{\beta_\mathrm{RS} c} = t_\mathrm{ej,f4} + \frac{\beta_1 t_\mathrm{off}}{\beta_4 - \beta_1} \left( 1 - \frac{\beta_4}{\beta_\mathrm{RS}} \right) 
\end{eqnarray}
The effective ejection timescales assume that the shells are ejected with post collision shock front velocities.  

The equations for the equal arrival time surfaces can be represented as
\begin{eqnarray}
    &\ \textit{EATS 1:}\hspace{2cm} \Delta T_\mathrm{RS} \equiv  \frac{T - T^\mathrm{eff}_\mathrm{ej,RS}}{(1+z)} = t - t^\mathrm{eff}_\mathrm{ej,RS} - \frac{R_\mathrm{RS} \cos \theta}{c} \\  \label{EATS1}
     &\ \textit{EATS  2:} \hspace{2cm} \Delta T_\mathrm{FS} \equiv  \frac{T - T^\mathrm{eff}_\mathrm{ej,FS}}{(1+z)} = t - t^\mathrm{eff}_\mathrm{ej,FS} - \frac{R_\mathrm{FS} \cos \theta }{c}  \\ \label{EATS2}
\end{eqnarray}

Both EATS intersect at $T=T_\mathrm{s}, t = t_\mathrm{o}, \theta = 0, R_\mathrm{FS}= R_\mathrm{RS} = R_\mathrm{o}$ which gives 
\begin{equation}
    \frac{T^\mathrm{eff}_\mathrm{ej,FS} - T^\mathrm{eff}_\mathrm{ej,RS}}{(1+z)} = t^\mathrm{eff}_\mathrm{ej,FS} - t^\mathrm{eff}_\mathrm{ej,RS} = \frac{\beta_1 \beta_4 t_\mathrm{off}}{(\beta_4 - \beta_1)} \left[ \frac{1}{\beta_\mathrm{RS}} - \frac{1}{\beta_\mathrm{FS}} \right] = \frac{R_\mathrm{o}}{c} \left[ \frac{1}{\beta_\mathrm{RS}} - \frac{1}{\beta_\mathrm{FS}} \right]
\end{equation}
which on further simplifcation can be expressed as
\begin{equation}
    T^\mathrm{eff}_\mathrm{ej,FS} =  T^\mathrm{eff}_\mathrm{ej,RS} + \frac{R_\mathrm{o} (1+z)}{c} \left[ \frac{1}{\beta_\mathrm{RS}} - \frac{1}{\beta_\mathrm{FS}} \right] \label{Tej}
\end{equation}

Besides, for $t = t_\mathrm{o}, \theta = 0, R_\mathrm{RS} = R_\mathrm{o}$, we have $T_\mathrm{s}- T^\mathrm{eff}_\mathrm{ej,RS} = T_\mathrm{0,RS}$ and $T_\mathrm{s} - T^\mathrm{eff}_\mathrm{ej,FS} = T_\mathrm{0,FS}$ which from equations (\ref{EATS1}) and (\ref{EATS2}) are given by
\begin{eqnarray}
\label{T0RS} &\    T_\mathrm{0,RS} = (1+z) \left( t_\mathrm{o} - t^\mathrm{eff}_\mathrm{ej,RS} - \frac{R_\mathrm{o}}{c}\right) = (1+z) \left( t_\mathrm{o} - t_\mathrm{o} + \frac{R_\mathrm{o}}{\beta_\mathrm{RS}c} - \frac{R_\mathrm{o}}{c}\right) = (1+z) \frac{R_\mathrm{o}}{c} \left( \frac{1 - \beta_\mathrm{RS}}{\beta_\mathrm{RS}} \right)  \\ 
&\    T_\mathrm{0,FS} = (1+z) \left( t_\mathrm{o} - t^\mathrm{eff}_\mathrm{ej,FS} - \frac{R_\mathrm{o}}{c}\right) = (1+z) \left( t_\mathrm{o} - t_\mathrm{o} + \frac{R_\mathrm{o}}{\beta_\mathrm{FS}c} - \frac{R_\mathrm{o}}{c}\right) = (1+z) \frac{R_\mathrm{o}}{c} \left( \frac{1 - \beta_\mathrm{FS}}{\beta_\mathrm{FS}} \right) \label{TOFS} 
\end{eqnarray}

We define $\widetilde{T}_\mathrm{RS}$ as
\begin{equation}
    \widetilde{T}_\mathrm{RS} = \frac{T - T^\mathrm{eff}_\mathrm{ej,RS}}{T_\mathrm{0,RS}} \label{tilde_TRS}
\end{equation}

Similarly we can define and simplify $\widetilde{T}_\mathrm{FS}$ using equations (\ref{T0RS})-(\ref{tilde_TRS}) as
\begin{equation}
    \widetilde{T}_\mathrm{FS} = \frac{T - T^\mathrm{eff}_\mathrm{ej,FS}}{T_\mathrm{0,FS}} = \frac{T_\mathrm{0,RS}}{T_\mathrm{0,FS}} \left[ \frac{T - T^\mathrm{eff}_\mathrm{ej,FS}}{T_\mathrm{0,RS}} \right] = \frac{T_\mathrm{0,RS}}{T_\mathrm{0,FS}} \left[ \frac{T - T^\mathrm{eff}_\mathrm{ej,RS}}{T_\mathrm{0,RS}} - \frac{R_\mathrm{o}(1+z)}{T_\mathrm{0,RS}c} \left( \frac{1}{\beta_\mathrm{RS}}- \frac{1}{\beta_\mathrm{FS}} \right) \right]  = \frac{\beta_\mathrm{FS}}{\beta_\mathrm{RS}} \left( \frac{1 - \beta_\mathrm{RS}}{1 - \beta_\mathrm{FS}} \right) \left[ \widetilde{T}_\mathrm{RS} - \frac{(\beta_\mathrm{FS} - \beta_\mathrm{RS})}{\beta_\mathrm{FS} (1 - \beta_\mathrm{RS})} \right] \label{tilde_TFS}
\end{equation}

In particular for $\widetilde{T}_\mathrm{FS}$ we have
\begin{equation}
    \widetilde{T}_\mathrm{FS} \lvert_{\widetilde{T}_\mathrm{RS} = 1} = \frac{\beta_\mathrm{FS}}{\beta_\mathrm{RS}} \left( \frac{1 - \beta_\mathrm{RS}}{1 - \beta_\mathrm{FS}} \right) \left[ \frac{\beta_\mathrm{FS} (1 - \beta_\mathrm{RS}) - \beta_\mathrm{FS} + \beta_\mathrm{RS}}{\beta_\mathrm{FS} (1 - \beta_\mathrm{RS})} \right] = 1, 
\end{equation}
as it should be for the first photons from the shocked regions must arrive simultaneously. 

We can also define 
\begin{equation}
    \overline{T}_\mathrm{RS} = \widetilde{T}_\mathrm{RS} - 1 \hspace{2cm} ; \hspace{2cm} \overline{T}_\mathrm{FS} = \widetilde{T}_\mathrm{FS} - 1 \label{bar_TRSFS}
\end{equation}
which when substituted in \ref{tilde_TFS} gives us,
\begin{equation}
    \overline{T}_\mathrm{FS} = \frac{T_\mathrm{0,RS}}{T_\mathrm{0,FS}} \; \overline{T}_\mathrm{RS} = \frac{\beta_\mathrm{FS}}{\beta_\mathrm{RS}} \left( \frac{1 - \beta_\mathrm{RS}}{1 - \beta_\mathrm{FS}} \right) \overline{T}_\mathrm{RS}= \left(\frac{\Gamma_\mathrm{FS}}{\Gamma_\mathrm{RS}} \right)^2 \frac{\beta_\mathrm{FS}}{\beta_\mathrm{RS}} \left( \frac{1 + \beta_\mathrm{FS}}{1 + \beta_\mathrm{RS}} \right) \overline{T}_\mathrm{RS} , 
\end{equation}

\section{ Estimating $F_{\nu}$ from a given shock front
}\label{appH}

\begin{table*}
    \centering
    \caption{Symbols $\&$ definitions for hydrodynamical parameters associated with shock fronts.}
    \begin{tabular}{cc} \hline 
         Symbols   & Definition \\ \hline 
        $R_\mathrm{o}$  & The collision radii (which is same) for both shock fronts \\ 
        $R_\mathrm{f,i}$ & The final location of shock front $i$\\
        $\Gamma$ & The Lorentz factor of the shocked material (same for both regions) \\
        $\Gamma_\mathrm{i}$ & The Lorentz factor of the shock front $i$ \\
        $R_\mathrm{L,i}$ & The largest radius on the equal arrival time surface at at an observer time $T$ for shock front $i$ \\
         \hline 
    \end{tabular}
    \label{hydro_param}
\end{table*}

Our treatment is a more refined version of the approach presented in \cite{2009MNRAS.399.1328G} (hereafter GG09). In our treatment, we distinguish between the LF of the shocked material and the shock front. Besides, we estimate the final location of each of the shock fronts based on the underlying shock hydrodynamics and the activity timescale of the compact source.  For sake of convenience, throughout the derivation subscript $i$ is dropped for dynamical quantities for a given shock front $i$. In the main text, all expressions use the subscript $i$ for quantities corresponding to shock fronts $i=(RS,FS)$ .    

Assuming that the emission is isotropic in the co-moving frame of the shocked fluid, the flux $F_{\nu} (T)$ received by an observer time at observed $T$ is given by
\begin{equation}
    \begin{split}
     F_{\nu} (T) &\ = \frac{1+z}{4 \pi d^2_\mathrm{L}} \int dL_{\nu}  
      =   \frac{1+z}{4 \pi d^2_\mathrm{L}} \int \frac{d \Omega}{4 \pi}  \; \delta^3 \; L'_{\nu'} 
      =  \frac{1+z}{4 \pi d^2_\mathrm{L}} \int^{2 \pi}_{0} \frac{d\phi}{4 \pi} \; \int d(\cos \theta) \; \delta^3 \; L'_{\nu'}  \\ &\ 
    = \frac{1+z}{8 \pi d^2_\mathrm{L}} \int dR \; \Bigg\lvert \frac{d (\cos \theta)}{dR} \Bigg \rvert \delta^3 L'_{\nu'} 
 = \frac{1+z}{8 \pi d^2_\mathrm{L}} \int^{y_\mathrm{max}}_{y_\mathrm{min}} dy \; \Bigg\lvert \frac{d \mu}{dy}  \Bigg \rvert \delta^3 L'_{\nu'}     
    \end{split} \label{Flux_general}
\end{equation}
where the limits $(y_\mathrm{min},y_\mathrm{max})$ are defined in eqns. (F32)-(F33) and the quantities $\mu$ and $\delta$ can be expressed as
\begin{eqnarray}
 \label{mu}   &\ \mu \equiv \cos \theta \\ 
    &\ \delta = (1+z) \frac{\nu}{\nu'} = \frac{1}{\Gamma (1 - \beta \mu)} \label{delta} 
\end{eqnarray}
We introduce the dimensionless variable $y$ as
\begin{equation}
    y = \frac{R}{R_\mathrm{L}}
\end{equation}
where $R_\mathrm{L}$ is the largest radius on the EATS at a given observer time $T$. 

We will derive the generic expression for \ref{Flux_general} for a radial dependence $(m \neq 0)$ of the shock front Lorentz factor $\Gamma^2_\mathrm{sh} = \Gamma^2_\mathrm{o} (R/R_\mathrm{o})^{-m}$ and then in accordance with our requirement consider the particular case of $m=0$ (where the Lorentz factor of the shock front is independent of the radial location).  Besides, we will assume that the shock front is ultra-relativistic ($\Gamma_\mathrm{sh} \gg 1$) and the emission originates from very small opening angles ($\theta \ll 1$).  

The Lorentz factor of the shock front $\Gamma_\mathrm{sh}$ can be expressed as 
\begin{equation}
\Gamma_\mathrm{sh} = \Gamma_\mathrm{o} \left( \frac{R}{R_\mathrm{o}}\right)^{-m/2}   =   \Gamma_\mathrm{o} \left( \frac{y R_\mathrm{L}}{R_\mathrm{o}}\right)^{-m/2} =  \Gamma_\mathrm{L} y^{-m/2}
\end{equation}
where $\Gamma_\mathrm{o}$ is the Lorentz factor of the shock front at the collision radius $R_\mathrm{o}$ and $\Gamma_\mathrm{L} = \Gamma_\mathrm{o} (R_\mathrm{L}/R_\mathrm{o})^{-m/2}$.

The Lorentz factor of the shocked material $\Gamma$ can be expressed as
\begin{equation}
    \Gamma = g \Gamma_\mathrm{sh} = g \Gamma_\mathrm{L} y^{-m/2} \label{gamma_m}
\end{equation}

The lab frame time $t$ is given as 
\begin{equation}
    t - t^\mathrm{eff}_\mathrm{ej} = \int^{R}_{0} \frac{dR}{\beta_\mathrm{sh} c} = \frac{1}{c} \int^{R}_{0} dR \left( 1 - \frac{1}{\Gamma^2_\mathrm{sh} }\right)^{-1/2}  \approx \frac{1}{c} \int^{R}_{0} dR \left( 1 + \frac{1}{2 \Gamma^2_\mathrm{sh} }\right) = \frac{R}{c} + \frac{R}{2(m+1)c \Gamma^2_\mathrm{sh}}    \hspace{2cm} \text{For $\Gamma_\mathrm{sh} \gg 1$}
\end{equation}

The equation of EATs is given as
\begin{equation}
    \frac{T - T^\mathrm{eff}_\mathrm{ej}}{(1+z)} = t - t^\mathrm{eff}_\mathrm{ej} - \frac{R \cos \theta}{c} \approx \frac{R}{c} \left[ 1 - \cos \theta + \frac{1}{2(m+1) \Gamma^2_\mathrm{sh}} \right] \approx \frac{R}{2c} \left[ \theta^2 + \frac{1}{(m+1)\Gamma^2_\mathrm{sh}}\right] = \frac{y R_\mathrm{L}}{2c} \left[ \theta^2 + \frac{1}{(m+1)\Gamma^2_\mathrm{sh}}\right] \hspace{1cm} \text{For $\Gamma_\mathrm{sh} \gg 1$ and $\theta \ll 1$} \label{EATs}
\end{equation}

The largest radius $R_\mathrm{L}$ at any observer time $T$ on EATs can be obtained by putting $(\theta=0, y = 1)$ in equation (\ref{EATs}), 
\begin{equation}
    R_\mathrm{L} = 2 (m+1) c \Gamma^2_\mathrm{L} \frac{T - T^\mathrm{eff}_\mathrm{ej}}{(1+z)}  \label{RL}
\end{equation}

Substituting equation (\ref{RL}) in equation (\ref{EATs}) we obtain
\begin{equation}
    \theta^2 = \frac{1}{(m+1)\Gamma^2_\mathrm{L}} (y^{-1} - y^{m}) \label{theta} 
\end{equation}

Using equation (\ref{theta}) we obtain,
\begin{equation}
    1 - \mu \equiv 1 - \cos \theta \approx 1 - (1-\frac{\theta^2}{2}) = \frac{\theta^2}{2} = \frac{1}{2(m+1)\Gamma^2_\mathrm{L}} (y^{-1} - y^{m}) \hspace{1cm}; \hspace{1cm} \frac{d\mu}{dy} = \frac{1}{2(m+1)\Gamma^2_\mathrm{L}} (y^{-2} + m y^{m-1}) \label{deri_mu}
\end{equation}

The collision radius $R_\mathrm{o}$ can be obtained by putting $(\theta = 0, \Gamma_\mathrm{sh} = \Gamma_\mathrm{o}, T = T_\mathrm{s}, T_\mathrm{s} - T^\mathrm{eff}_\mathrm{ej} = T_{0} )$ in equation (\ref{EATs})
\begin{equation}
    R_\mathrm{o} = 2 (m+1) c \Gamma^2_\mathrm{o} \frac{T_\mathrm{0}}{1+z} \label{R0}
\end{equation}

Using equation (\ref{RL}) and equation (\ref{R0}) we have,
\begin{equation}
    \frac{R_\mathrm{L}}{R_\mathrm{o}} = \frac{\Gamma^2_\mathrm{L}}{\Gamma^2_\mathrm{o}} \left( \frac{T - T^\mathrm{eff}_\mathrm{ej}}{T_\mathrm{0}} \right) = \frac{\Gamma^2_\mathrm{L}}{\Gamma^2_\mathrm{o}} \widetilde{T}  \hspace{2cm} \text{where $\widetilde{T} = \frac{T - T^\mathrm{eff}_\mathrm{ej}}{T_\mathrm{0}} $} \label{ratio}
\end{equation}

Using equation (\ref{ratio}) in the definition of $\Gamma_\mathrm{L}$ we have,
\begin{equation}
    \Gamma_\mathrm{L} = \Gamma_\mathrm{o} \left( \frac{R_\mathrm{L}}{R_\mathrm{o}} \right)^{-m/2} = \Gamma_\mathrm{o} \left( \frac{\Gamma_\mathrm{L}}{\Gamma_\mathrm{o}}\right)^{-m} \widetilde{T}^{-\frac{m}{2} } \Rightarrow \Gamma_\mathrm{L} = \Gamma_\mathrm{o} \widetilde{T}^{- \frac{m}{2(m+1)} } \label{GammaL}
\end{equation}

Using equation (\ref{gamma_m}) and equation (\ref{theta}) we have in the ultra-relativistic approximation
\begin{equation}
    1 - \beta \mu \approx (1 - \beta ) + (1- \mu) = \frac{1}{2 \Gamma^2 } + \frac{\theta^2}{2} = \frac{1}{2} \left[ \frac{1}{g^2 \Gamma^2_\mathrm{L} y^{-m}} + \frac{y^{-1} - y^{m}}{(m+1)\Gamma^2_\mathrm{L}}  \right] = \frac{(m+1)y^{m} + g^2(y^{-1} - y^{m})}{2(m+1)g^2 \Gamma^2_\mathrm{L}} \label{approx1}
\end{equation}

Using equation (\ref{gamma_m}) and (\ref{approx1}) in equation (\ref{delta}) we have
\begin{equation}
    \delta = (1+z) \frac{\nu}{\nu'} = \frac{1}{\Gamma ( 1 - \beta \mu)} \approx \frac{1}{g \Gamma_\mathrm{L} y^{-\frac{m}{2}}} \left[ \frac{2(m+1)g^2 \Gamma^2_\mathrm{L}}{(m+1)y^{m} + g^2(y^{-1} - y^{m})} \right] = \frac{2(m+1)g \Gamma_\mathrm{L} y^{\frac{m}{2}}}{(m+1)y^{m} + g^2(y^{-1} - y^{m})} \label{doppler}
\end{equation}

Using equation (\ref{deri_mu}) and equation (\ref{doppler}) in equation (\ref{Flux_general}) we get
\begin{equation}
    F_{\nu} = \frac{g^3 (1+z) \Gamma_\mathrm{L}}{2 \pi d^2_\mathrm{L}} \int^{y_\mathrm{max}}_{y_\mathrm{min}} dy \; y^{-1 - \frac{m}{2}} \left[ \frac{m+1}{m+1+g^2(y^{-1-m} - 1)} \right]^2 \left( \frac{y^{-2} + m y^{m-1}}{g^2 y^{-2} + (1-g^2+m) y^{m-1}} \right) L'_{\nu'} \label{F_nu_expr1}
\end{equation}

Using equation (\ref{GammaL}) in equation (\ref{F_nu_expr1}) we get,
\begin{equation}
    F_{\nu} = \frac{g^3 (1+z) \Gamma_\mathrm{o} \widetilde{T}^{- \frac{m}{2(m+1)} }  }{2 \pi d^2_\mathrm{L}} \int^{y_\mathrm{max}}_{y_\mathrm{min}} dy \; y^{-1 - \frac{m}{2}} \left[ \frac{m+1}{m+1+g^2(y^{-1-m} - 1)} \right]^2 \left( \frac{y^{-2} + m y^{m-1}}{g^2 y^{-2} + (1-g^2+m) y^{m-1}} \right) L'_{\nu'} 
\end{equation}

For $m=0$ we obtain the particular case we are interested in as
\begin{equation}
    F_{\nu} = \frac{g^3 (1+z) \Gamma_\mathrm{sh}}{2 \pi d^2_\mathrm{L}} \int^{y_\mathrm{max}}_{y_\mathrm{min}} \; dy\;  \frac{y^{-2}}{[1+g^2(y^{-1} - 1)]^3} \; L'_{\nu'} \label{F_nu_expr2} 
\end{equation}

Now, the comoving luminosity $L'_{\nu'}$ is given as
\begin{equation}
    L'_{\nu'} = L'_\mathrm{o} \left( \frac{R}{R_\mathrm{o}} \right)^{a} S\left[ \frac{\nu'}{\nu'_\mathrm{p}} \right]  = L'_\mathrm{o} \left( \frac{R_\mathrm{L}}{R_\mathrm{o}}\right)^{a} y^{a} \; S\left[ \frac{\nu'}{\nu'_\mathrm{p}} \right]  = L'_\mathrm{o} \widetilde{T}^{a} \; y^{a} \; S\left[ \frac{\nu'}{\nu'_\mathrm{p}} \right]  \label{Lnu}
\end{equation}
where
\begin{eqnarray}
    &\ \nu' = \frac{(1+z) \nu}{\delta} = (1+z) \nu  \frac{[1+g^2(y^{-1} - 1)]}{2g \Gamma_\mathrm{sh}} = (1+z) \nu  \frac{[1+g^2(y^{-1} - 1)]}{2\Gamma} \\
    &\ \nu'_\mathrm{p} = \nu'_\mathrm{o} \left( \frac{R}{R_\mathrm{o}}\right)^{d} = (1+z) \nu_\mathrm{o} \;  \frac{1}{2 \Gamma} \left( \frac{y R_\mathrm{L}}{R_\mathrm{o}} \right)^{d} =   (1+z) \nu_\mathrm{o} \;  \frac{1}{2 \Gamma} \; y^{d} \; \widetilde{T}^{d}
\end{eqnarray}

Substituting equation (\ref{Lnu}) in equation (\ref{F_nu_expr2}) we get,
\begin{equation}
F_{\nu} =    \frac{g^3 L'_\mathrm{o} \; (1+z) \Gamma_\mathrm{sh}}{2 \pi d^2_\mathrm{L}} \; \widetilde{T}^{a} \int^{y_\mathrm{max}}_{y_\mathrm{min}} \; dy\;  \frac{y^{a-2}}{[1+g^2(y^{-1} - 1)]^3} \; S\left[ \frac{\nu'}{\nu'_\mathrm{p}} \right]
\end{equation}
where 
\begin{equation}
 \frac{\nu'}{\nu'_\mathrm{p}} = \frac{\nu}{\nu_\mathrm{o}} \;  \widetilde{T}^{-d} \; y^{-d}  [1 + g^2(y^{-1} - 1)] = \widetilde{\nu} \;  \widetilde{T}^{-d} \; y^{-d}  [1 + g^2(y^{-1} - 1)]
\end{equation}
where we define $\widetilde{\nu} = \nu/\nu_\mathrm{o}$.

For the particular case of internal shock model we are interested in we have $(a=1,d=-1)$ such that
\begin{equation}
    F_{\nu} =    \frac{g^3 L'_\mathrm{o} \; (1+z) \Gamma_\mathrm{sh}}{2 \pi d^2_\mathrm{L}} \; \widetilde{T} \int^{y_\mathrm{max}}_{y_\mathrm{min}} \; dy\;  \frac{y^{-1}}{[1+g^2(y^{-1} - 1)]^3} \; S\left[ x \right] \label{F_nu_expr3}
\end{equation}
where
\begin{equation}
    x =  \frac{\nu'}{\nu'_\mathrm{p}}\Bigg\lvert_{d=-1} =  (\widetilde{\nu} \; \widetilde{T} ) \; y [1+g^2(y^{-1} - 1)] = \widetilde{A}  y [1+g^2(y^{-1} - 1)]
\end{equation}
where we define $\widetilde{A} = \widetilde{\nu} \; \widetilde{T}$.

Next, we change the variable of integration from $y$ to x such that we have for $(g \neq 1)$
\begin{eqnarray}
&\ y = \frac{\widetilde{A} x^{-1} - g^2}{1-g^2} \\ 
&\ dy = \frac{\widetilde{A}^{-1}}{1-g^2} dx  \\ 
\end{eqnarray}
which when substituted in equation (\ref{F_nu_expr3}) gives
\begin{equation}
    F_{\nu} = \left( \frac{2 g \Gamma_\mathrm{sh} (1+z) L'_\mathrm{o}}{12 \pi d^2_{L}} \right) \frac{ 3 g^2}{(1-g^2)^3}  \widetilde{A}^2 \; \widetilde{T} \int^{x_\mathrm{max}}_{x_\mathrm{min}} \; dx \; x^{-3} (\widetilde{A}^{-1} x - g^2)^2 \; S[x] = F_\mathrm{o} \; \frac{3 g^2}{(1-g^2)^3}  \widetilde{\nu}^2 \; \widetilde{T}^3 \int^{x_\mathrm{max}}_{x_\mathrm{min}} \; dx \; x^{-3} (\widetilde{A}^{-1} x - g^2)^2 \; S[x] \label{F_nu_expr4}
\end{equation}
where $F_\mathrm{o} = 2 g \Gamma_\mathrm{sh} (1+z) L'_\mathrm{o}/12 \pi d^2_{L} = 2 \Gamma (1+z) L'_\mathrm{o}/12 \pi d^2_{L}$ and the integration limits are
\begin{equation}
    x_\mathrm{min} = \widetilde{A} [1 + g^2(y^{-1}_\mathrm{min} - 1)]  \hspace{2cm} \; \hspace{2cm} x_\mathrm{max} = \widetilde{A} [1 + g^2(y^{-1}_\mathrm{max} - 1)]  
\end{equation}
where for the $i$-th shock front we have
\begin{eqnarray}
&\    y_\mathrm{min} = \min\left[ 1, \frac{R_\mathrm{o}}{R_\mathrm{L}(T)} \right] \begin{cases}
       &\ 1    \hspace{2cm} \text{For $T \leq T^\mathrm{eff}_\mathrm{ej,i}$} + T_\mathrm{0,i} \\
       &\ \left( \frac{T - T^\mathrm{eff}_\mathrm{ej,i}}{T_\mathrm{0,i}} \right)^{-1} \hspace{0.8cm} \text{For $T \geq T^\mathrm{eff}_\mathrm{ej,i}$} + T_\mathrm{0,i}  
    \end{cases} \\ 
    &\    y_\mathrm{max} = \min\left[ 1, \frac{R_\mathrm{o} + \Delta R_\mathrm{i}}{R_\mathrm{L}(T)} \right] \begin{cases}
       &\ 1    \hspace{2cm} \text{For $T \leq T^\mathrm{eff}_\mathrm{ej,i}$} + T_\mathrm{f,i} \\
       &\ \left( \frac{T - T^\mathrm{eff}_\mathrm{ej,i}}{T_\mathrm{f,i}} \right)^{-1} \hspace{0.8cm} \text{For $T \geq T^\mathrm{eff}_\mathrm{ej,i}$} + T_\mathrm{f,i} 
      \end{cases} 
\end{eqnarray}
where $T_\mathrm{f,i} = T_{0,i} \left( 1 + \frac{\Delta R}{R_\mathrm{o}}\right) $. 

We assume the following functional form for $S[x]$ as
\begin{equation}
  S[x] = e^{1+b_1} 
  \begin{cases}
      &\ x^{b_{1}} e^{-(1+b_1) x}  \hspace{2cm} \text{$x\leq x_\mathrm{b}$} \\
      &\ x^{b_{2}} x^{b_1 - b_2}_\mathrm{b} e^{-(b_1 - b_2)} \hspace{1.1cm}  \text{$x\geq x_\mathrm{b}$}
  \end{cases} \label{Sx2}
\end{equation}

Substituting equation (\ref{Sx2}) in equation (\ref{F_nu_expr4}) we get,
\begin{equation}
    \frac{F_{\nu}}{F_\mathrm{o}} =  \frac{3 g^2}{(1-g^2)^3} \widetilde{\nu}^2 \; \widetilde{T}^3 
    \begin{cases}
        &\ e^{1+b_1} \int^{x_\mathrm{max}}_{x_\mathrm{min}} \; dx \; e^{-(1+b_1)x} \; x^{-3 + b_1} \;  (\widetilde{A}^{-1} x - g^2)^2 \hspace{6.8cm} \text{$(x_\mathrm{min},x_\mathrm{max}) < x_\mathrm{b}$} \\
        &\ x^{b_1 - b_2}_\mathrm{b} \; e^{1+b_2} \; \int^{x_\mathrm{max}}_{x_\mathrm{min}} \; dx \; \; x^{-3 + b_2} \;  (\widetilde{A}^{-1} x - g^2)^2 \hspace{7cm} \text{$(x_\mathrm{min},x_\mathrm{max}) >  x_\mathrm{b}$}  \\
        &\  e^{1+b_1} \int^{x_\mathrm{b}}_{x_\mathrm{min}} \; dx \; e^{-(1+b_1)x} \; x^{-3 + b_1} \;  (\widetilde{A}^{-1} x - g^2)^2 + x^{b_1 - b_2}_\mathrm{b} \; e^{1+b_2} \; \int^{x_\mathrm{max}}_{x_\mathrm{b}} \; dx \; \; x^{-3 + b_2} \;  (\widetilde{A}^{-1} x - g^2)^2 \hspace{0.5cm}\text{ $x_\mathrm{min} < x_\mathrm{b}< x_\mathrm{max}$}  
    \end{cases}
\end{equation}

The integrals have analytical solution as 
\begin{eqnarray}
    \begin{split}
       &\ \int \; dx \; e^{-(1+b_1)x} \; x^{-3 + b_1} (\widetilde{A}^{-1} x - g^2)^2  \\
       &\ = - \frac{x^{b_1} ((b_1+1)x)^{-b_1}}{\widetilde{A}^2} \left[ \widetilde{A}^2 (b_1+1)^2 g^4 \Gamma(b_1-2,(b_1+1)x) - 2 \widetilde{A} (b_1 + 1) g^2 \Gamma(b_1-1,(b_1+1)x) + \Gamma(b_1,(b_1+1)x)  \right] 
    \end{split} \\ 
    \int \; dx \; \; x^{-3 + b_2} \; (\widetilde{A}^{-1} x - g^2)^2 = \frac{x^{b_2}}{\widetilde{A}^2} \left[ \widetilde{A} g^2 \left( \frac{\widetilde{A} g^2}{(b_2 -2)x^2} + \frac{2}{(1-b_2 )x}\right) + \frac{1}{b_2} \right]
\end{eqnarray}
where $\Gamma(b_1-2,(b_1+1)x)$, $\Gamma(b_1-1,(b_1+1)x)$ and $\Gamma(b_1,(b_1+1)x)$ are incomplete gamma function.

The normalized time $\overline{T}_\mathrm{f,i}$ corresponding to the final location of the shock front $i$ is given by 
\begin{equation}
    \overline{T}_\mathrm{f,i} =   
    \begin{cases}
        &\   \frac{\beta_\mathrm{RS} c t_\mathrm{RS}}{R_\mathrm{o}} = \left[ \left( \frac{\beta_4 - \beta_1}{\beta_4 - \beta_\mathrm{RS}} \right) \left( \frac{\beta_\mathrm{RS}}{\beta_1}\right)  \right] \left( \frac{t_\mathrm{on4}}{t_\mathrm{off}} \right) \hspace{5cm} \text{For $i = RS$} \\ 
        &\ \frac{\beta_\mathrm{RS} (1+\beta_\mathrm{RS})}{\beta_\mathrm{FS} (1+\beta_\mathrm{FS})} \left( \frac{\Gamma_\mathrm{RS}}{\Gamma_\mathrm{FS}} \right)^2 \frac{\beta_\mathrm{FS} c t_\mathrm{FS}}{R_\mathrm{o}} = \left[ \frac{\beta_\mathrm{RS} (1+\beta_\mathrm{RS}) (\beta_4 - \beta_1)}{\beta_\mathrm{1} (1+\beta_\mathrm{FS}) (\beta_\mathrm{FS} - \beta_1)} \left( \frac{\Gamma_\mathrm{RS}}{\Gamma_\mathrm{FS}} \right)^2 \right] \left( \frac{t_\mathrm{on1}}{t_\mathrm{off}} \right) \hspace{1.4cm}  \text{For $i = FS$}
    \end{cases}
\end{equation}

\section{Normalization constants}\label{appI}

In the main text all frequencies, flux, and times are normalized to $\nu_\mathrm{o} \equiv \nu_\mathrm{o,RS}$, $F_\mathrm{o} \equiv F_\mathrm{o,RS}$  and $T_\mathrm{0} \equiv T_\mathrm{0,RS}$. The objective of the present appendix is to give some dimensional values of the same using some fiducial values of the basic parameters of our model. As with the main text, we assume a constant source power $L$ during shell ejection. 

The collision radius is given by
\begin{equation}
    R_\mathrm{o} = \frac{\beta_1 \beta_4 c t_\mathrm{off}}{(\beta_4 - \beta_1)} \approx 2 \frac{\mathrm{u}^2_1 c t_\mathrm{off}}{(1 - a^{-2}_\mathrm{u})}
\end{equation}

The lab frame density of shell S4 just before the collision is given as
\begin{equation}
    \rho_{4,0} = \frac{M_{4}}{V_{4}} = \frac{M_4}{4 \pi R^2_\mathrm{o} \Delta_4} = \frac{1}{c^2}\left( \frac{E_\mathrm{k,4}}{\Gamma_4 -1} \right) \frac{1}{4 \pi R^2_\mathrm{o} \Delta_4} = \frac{1}{c^2}\left( \frac{L t_\mathrm{on4}}{\Gamma_4 -1} \right) \frac{1}{4 \pi R^2_\mathrm{o} } \frac{1}{\beta_4 c t_\mathrm{on4}} \approx \frac{L}{\Gamma_4} \frac{1}{4 \pi R^2_\mathrm{o}} \frac{1}{c^3}  \quad \quad \text{For $\Gamma_4 \gg 1$}
\end{equation}

The comoving density of shell S4 just before the collision is given as
\begin{equation}
    \rho'_{4,0} = \frac{\rho_4 }{\Gamma_4} \approx \frac{L}{\Gamma^2_4} \frac{1}{4 \pi R^2_\mathrm{o}} \frac{1}{c^3}  \quad \quad \text{For $\Gamma_4 \gg 1$} 
\end{equation}

The internal energy density in region R3 is given as
\begin{equation}
    e'_\mathrm{int3} = (\Gamma_{34} - 1) \rho'_{3,0} c^2 = 4 \Gamma_{34}  (\Gamma_{34} - 1) \rho'_{4,0} c^2 =  4 \Gamma_{34}  (\Gamma_{34} - 1) \left( \frac{L}{\Gamma^2_4} \frac{1}{4 \pi R^2_\mathrm{o}} \frac{1}{c}\right) 
\end{equation}
where the relative LF $\Gamma_{34}$ in the relativistic limit can be estimated as
\begin{equation}
    \Gamma_{34} \approx \frac{1}{2} \left( \frac{\Gamma_4}{\Gamma}+ \frac{\Gamma}{\Gamma_4} \right) = \frac{1}{2 \sqrt{2}} \left( \frac{3 + a^2_\mathrm{u}}{\sqrt{1 + a^2_\mathrm{u}}}\right) 
\end{equation}

Using equation (G4) the comoving magnetic field strength $B'$ can be estimated as
\begin{equation}
    B' = \sqrt{8 \pi \epsilon_\mathrm{B} e'_\mathrm{int3}} = \sqrt{32 \pi \epsilon_\mathrm{B} \Gamma_{34} (\Gamma_{34} -1 ) \left( \frac{L}{\Gamma^2_4} \frac{1}{4 \pi R^2_\mathrm{o}} \frac{1}{c}\right)  }  \approx \frac{\sqrt{2}}{c^{3/2}} \epsilon^{1/2}_\mathrm{B} \left[ \Gamma^{1/2}_{34} (\Gamma_{34}-1)^{1/2} (a^{-1}_\mathrm{u} - a^{-3}_\mathrm{u}) \right]  u^{-3}_{1} L^{1/2} t^{-1}_\mathrm{off}
\end{equation}

The minimal LF of the non-thermal distribution (for $p>2$) is given as
\begin{equation}
    \gamma_\mathrm{m} = \frac{m_\mathrm{p}}{m_\mathrm{e}} G(p) \frac{\epsilon_\mathrm{e}}{\xi_\mathrm{e}}  (\Gamma_{34} - 1 )
\end{equation}
where $G(p) = \frac{p-2}{p-1}$. 

The comoving cyclotron frequency is given as
\begin{equation}
    \nu'_\mathrm{B} = \frac{q_\mathrm{e} B' }{2 \pi m_\mathrm{e} c} \approx  \frac{q_{e}}{ \sqrt{2} \pi m_\mathrm{e} \; c^{5/2}} \epsilon^{1/2}_\mathrm{B} \left[ \Gamma^{1/2}_{34} (\Gamma_{34}-1)^{1/2} (a^{-1}_\mathrm{u} - a^{-3}_\mathrm{u}) \right]  u^{-3}_{1} L^{1/2} t^{-1}_\mathrm{off} 
\end{equation}

The peak frequency in the comoving frequency is given as 
\begin{equation}
    \nu'_\mathrm{o} = \gamma^2_\mathrm{m} \nu'_\mathrm{B} = \frac{1}{ \sqrt{2} \pi }  \frac{q_\mathrm{e} m^2_\mathrm{p}}{m^3_\mathrm{e} c^{5/2}} G^2(p) \xi^{-2}_\mathrm{e} \epsilon^2_\mathrm{e} \epsilon^{1/2}_\mathrm{B} \Gamma^{1/2}_{34} (\Gamma_{34} - 1)^{5/2} u_1^{-3} (a^{-1}_\mathrm{u} - a^{-3}_\mathrm{u}) L^{1/2} t^{-1}_\mathrm{off}
\end{equation}

The peak energy of the observed photons originating at $(R_\mathrm{o},\theta = 0)$ is given as
\begin{equation}
 \nu_\mathrm{o,RS}  =  \frac{2 \Gamma}{\left(1+z\right)}  \nu'_\mathrm{o} \approx \frac{1}{1+z} \frac{ 2 \sqrt{2} a_\mathrm{u} }{\sqrt{1+ a^2_\mathrm{u}}} u_{1} \nu'_\mathrm{o}
\end{equation}

The normalization frequency is given as
\begin{equation}
    \nu_\mathrm{o} = \nu_\mathrm{o,RS}= \frac{1}{\left( 1+z \right) } \left[ \frac{2}{ \pi } \frac{ q_\mathrm{e} m^2_\mathrm{p}}{m^3_\mathrm{e} c^{5/2}} \right] G^2(p) \xi^{-2}_\mathrm{e} \epsilon^2_\mathrm{e} \epsilon^{1/2}_\mathrm{B} \left[ \Gamma^{1/2}_{34} (\Gamma_{34} - 1)^{5/2}   \frac{(1 - a^{-2}_\mathrm{u})}{\sqrt{1 + a^2_\mathrm{u}}} \right] \; u_1^{-2}\; L^{1/2} \; t^{-1}_\mathrm{off} \label{norm_freq}
\end{equation}

From Appendix D, we have
\begin{equation}
    W(p) \nu'_\mathrm{o} L'_\mathrm{o} = L'_\mathrm{bol,\gamma} = \frac{4}{3} \epsilon_\mathrm{e} \epsilon_\mathrm{rad} (\Gamma_{34} - 1)u_{34} \left[ \rho'_{4,0} c^3 4 \pi R^2_\mathrm{o} \right]  \approx  \frac{4}{3} \epsilon_\mathrm{e} \epsilon_\mathrm{rad} (\Gamma_{34} - 1) u_{34} \frac{L}{\Gamma^2_4}
\end{equation}
where $W(p) = 2 \left( \frac{p-1}{p-2} \right) = \frac{2}{G(p)}$.

After some algebra, the comoving peak luminosity $L'_\mathrm{o}$ can be represented as 
\begin{equation}
    L'_\mathrm{o} =  \left[\frac{ 2\sqrt{2} \pi  }{3}  \frac{1}{q_\mathrm{e}} \frac{m^3_\mathrm{e}}{m^2_\mathrm{p}}  c^{5/2} \right] G^{-1}(p) \epsilon^{-1/2}_\mathrm{B} \epsilon^{-1}_\mathrm{e} \xi^{2}_\mathrm{e} \beta_{34} \Gamma^{1/2}_{34} (\Gamma_{34} - 1)^{-3/2} u_{1} (a_\mathrm{u} - a^{-1}_\mathrm{u})^{-1}  L^{1/2} t_\mathrm{off}
\end{equation}

The isotropic spectral flux in the observer frame is given as 
\begin{equation}
    F_\mathrm{s} = \frac{2 (1+z) \Gamma L'_\mathrm{o}}{4 \pi d^2_\mathrm{L}} = \frac{1+z}{4 \pi d^2_\mathrm{L}}  \left[\frac{8 \pi  }{3}  \frac{1}{q_\mathrm{e}} \frac{m^3_\mathrm{e}}{m^2_\mathrm{p}}  c^{5/2} \right] G^{-1}(p) \epsilon^{-1/2}_\mathrm{B} \epsilon^{-1}_\mathrm{e} \xi^{2}_\mathrm{e} \beta_{34} \Gamma^{1/2}_{34} (\Gamma_{34} - 1)^{-3/2} \frac{(1 - a^{-2}_\mathrm{u})^{-1}}{\sqrt{1 + a^2_\mathrm{u}}}  u^2_{1} L^{1/2} t_\mathrm{off}
\end{equation}

The flux normalization $F_\mathrm{o}$ is defined as
\begin{equation}
    F_\mathrm{o} = \frac{F_\mathrm{s}}{3} = \frac{(1+z)}{12 \pi d^2_\mathrm{L}}  \left[\frac{8 \pi }{3}  \frac{1}{q_\mathrm{e}} \frac{m^3_\mathrm{e}}{m^2_\mathrm{p}}  c^{5/2} \right] G^{-1}(p) \epsilon^{-1/2}_\mathrm{B} \epsilon^{-1}_\mathrm{e} \xi^{-2}_\mathrm{e} \left[ \beta_{34} \Gamma^{1/2}_{34} (\Gamma_{34} - 1)^{-3/2} \frac{(1 - a^{-2}_\mathrm{u})^{-1}}{\sqrt{1 + a^2_\mathrm{u}}} \right]   u^2_{1} L^{1/2} t_\mathrm{off} \label{flux_norm} 
\end{equation}

The time normalization constant $(T_{0} = T_\mathrm{0,RS})$ is given as 
\begin{equation}
    T_\mathrm{0,RS} = (1+z) \frac{R_\mathrm{o}}{c} \frac{1- \beta_\mathrm{RS}}{\beta_\mathrm{RS}} \approx (1+z) \frac{R_\mathrm{o}}{2 c \Gamma^2_\mathrm{RS}} =  (1+z) \frac{\Gamma^2_{1}}{4\Gamma^2_\mathrm{RS}} \frac{t_\mathrm{off}}{(1-a^{-2}_\mathrm{u})} =(1+z) \frac{1+a^2_\mathrm{u}}{4 a^2_\mathrm{u}} g^2_\mathrm{RS} \frac{1}{(1-a^{-2}_\mathrm{u})} t_\mathrm{off} = (1+z) \left( \frac{g_\mathrm{RS}}{2}\right)^2  \left(  \frac{a^2_\mathrm{u} + 1}{a^2_\mathrm{u} - 1} \right) t_\mathrm{off} \label{time_norm}
\end{equation}

Using equations (\ref{norm_freq}) and (\ref{flux_norm}) can estimate the product $\nu_\mathrm{o} F_\mathrm{o}$ as follows:
\begin{equation}
    \nu_\mathrm{o} F_\mathrm{o} = \frac{4} {9 \pi d^2_\mathrm{L}} G(p) \; \left[ \frac{1}{1+a^2_\mathrm{u}}  (\Gamma_\mathrm{34} - 1) u_\mathrm{34} \right] \epsilon_\mathrm{e} \epsilon_\mathrm{rad}  L  \label{norm_nuF}
\end{equation}

Equations (\ref{norm_freq}), and (\ref{time_norm}) and (\ref{norm_nuF}) can be expressed
as in terms of using fiducial values for the basic parameters
\begin{eqnarray}
    &\ h \nu_\mathrm{o} = 0.44 \; \left[ \frac{1 + z}{2} \right]^{-1} \; \left[ \frac{G(p)}{G(2.5)}\right]^2 \; \left[ \frac{F(a_\mathrm{u})}{F(2)} \right] \;\xi^{-2}_\mathrm{e,-2} \epsilon^2_\mathrm{e,-0.47} \epsilon^{1/2}_\mathrm{B,-0.47}\; \; u_{1,2}^{-2}\; L^{1/2}_\mathrm{53} \; t^{-1}_\mathrm{off,-1} \;  \text{MeV} \\
   &\ \nu_\mathrm{o} F_\mathrm{o} = 4 \times 10^{-8} \; d^{-2}_\mathrm{L,28.3} \left[ \frac{G(p)}{G(2.5)} \right] \; \left[\frac{H(a_\mathrm{u})}{H(2)}\right] \epsilon_\mathrm{e,-0.47} \epsilon_\mathrm{rad} \; L_\mathrm{53} \; \text{ergs s$^{-1}$ cm$^{-2}$} \\
   &\  T_\mathrm{0}= 0.22 \; \left[ \frac{1+z}{2} \right]  \left( \frac{g}{1.15}\right)^2  \left[ \frac{I(a_\mathrm{u})}{I(2)} \right] t_\mathrm{off,-1} \; \text{s} 
\end{eqnarray}
where we use the usual convention $Q_\mathrm{x} = Q/10^\mathrm{x}$ in cgs units, and where 
\begin{eqnarray}
&\    G(p)   = \frac{p-2}{p-1}   \\ 
&\ F(a_\mathrm{u}) =  \Gamma^{1/2}_{34} (\Gamma_{34} - 1)^{5/2}   \frac{(1 - a^{-2}_\mathrm{u})}{\sqrt{1 + a^2_\mathrm{u}}}    \\ &\ H(a_\mathrm{u}) =   \frac{1}{1+a^2_\mathrm{u}}  (\Gamma_\mathrm{34} - 1) u_\mathrm{34}   \\
&\  I(a_\mathrm{u}) = \frac{a^2_\mathrm{u} + 1}{a^2_\mathrm{u} - 1}             \\
&\  \Gamma_{34} \approx \frac{1}{2} \left( \frac{\Gamma_4}{\Gamma}+ \frac{\Gamma}{\Gamma_4} \right) = \frac{1}{2 \sqrt{2}} \left( \frac{3 + a^2_\mathrm{u}}{\sqrt{1 + a^2_\mathrm{u}}}\right) 
\end{eqnarray}

The canonical values of the various parameters used in the normalization expressed in Eqns. (G18) to (G20) are given as $\epsilon_\mathrm{B} = \epsilon_\mathrm{e} = \frac{1}{3},\xi_\mathrm{e} = 10^{-2}, 
p = 2.5, u_1 = 100, a_\mathrm{u} = 2, t_\mathrm{off} = 0.1 \;\text{s}, L = 10^{53} \hspace{0.1cm} \text{ergs/s}, z = 1, d_\mathrm{L} = 2 \times 10^{28} \hspace{0.1cm} \text{cm}$

\section{Estimating effective angular timescales}\label{appJ}

The Doppler factor $\delta$ for an ultra-relativistic flow $\Gamma \gg 1$ and small angle approximation $\theta \ll 1$ is given as
\begin{equation}
    \delta \approx \frac{2 \Gamma }{1 + (\Gamma \theta)^2}  \propto \left[ 1 + (\Gamma \theta)^2 \right]^{-1} \label{Doppler_dep}
\end{equation}
as the LF of the shocked material $\Gamma$ remains constant. 

The equation for EATS for $(\Gamma \gg 1,\theta\ll 1)$ can be written as
\begin{equation}
    \frac{T - T^\mathrm{eff}_\mathrm{ej}}{ (1+z)} = \left( t - t^\mathrm{eff}_\mathrm{ej} - \frac{R}{c} \right)  + \frac{R }{c} (1- \cos \theta) \approx \frac{R}{2 c \Gamma^2_\mathrm{sh}} + \frac{R}{2 c \Gamma^2 } (\Gamma \theta )^2
\end{equation}
which can be represented as
\begin{equation}
    T - T^\mathrm{eff}_\mathrm{ej} = T_\mathrm{R} + T_\mathrm{\theta} (\Gamma \theta)^2 \Rightarrow (\Gamma \theta)^2 = \frac{T - T^\mathrm{eff}_\mathrm{ej}-T_\mathrm{R}}{T_{\theta}} \label{EATS_ang}
\end{equation}
where $T_\mathrm{R}$ and $T_\mathrm{\theta}$ are the radial and the angular time at a distance $R$ from the source and expressed as 
\begin{equation}
    T_\mathrm{R} = \frac{R}{2 c \Gamma^2_\mathrm{sh}} = g^2  \left(\frac{R}{2 c \Gamma^2 } \right) = g^2 T_{\theta} 
\end{equation}

Using equation (\ref{EATS_ang}) we can define an effective angular timescale 
\begin{equation}
    \widetilde{T}_\mathrm{eff} = 1 + (\Gamma \theta)^2 = \left( 1 - \frac{T_\mathrm{R}}{T_{\theta}}\right) + \frac{T - T^\mathrm{eff}_\mathrm{ej}}{T_{\theta}} = (1 - g^2) + g^2 \left( \frac{T - T_\mathrm{ej}}{T_\mathrm{R}} \right) = (1 - g^2) + g^2 \widetilde{T}_\mathrm{R}
\end{equation}
In our notation,
\begin{equation}
    \widetilde{T} = \widetilde{T}_\mathrm{R} (R=R_\mathrm{o}) 
\end{equation}
Since the radial (and the angular) timescale linearly with distance in our analysis, we have
\begin{equation}
    T_\mathrm{R} = \frac{R}{R_\mathrm{o}} T_\mathrm{o}
\end{equation}

For our analysis, two effective angular times of special interest are for $R_\mathrm{o}$ and the final location $R_\mathrm{f,i}$ of the shock front $i$ expressed as 
\begin{eqnarray}
    &\ \widetilde{T}_\mathrm{eff1,i} = (1 - g^2_\mathrm{i}) + g^2_\mathrm{i} \widetilde{T}_\mathrm{i} \\
     &\ \widetilde{T}_\mathrm{eff2,i} = (1 - g^2_\mathrm{i}) + g^2_\mathrm{i} \left( \frac{R_\mathrm{o}}{R_\mathrm{f,i}} \right) \widetilde{T}_\mathrm{i}
\end{eqnarray}

\section{Estimating $\Delta R/R_\mathrm{o}$ for representative samples of GRB}\label{radial_est}

For collision of ultra-relativistic shells, the ratio of the radial extent of the reverse shocked region to the collision radii can be represented as  
\begin{equation}
    \frac{\Delta R}{R_\mathrm{o}} \approx \Bigg \{   2 \left( \frac{a^2_\mathrm{u} - 1}{a^2_\mathrm{u} + 1} \right)  \frac{1}{g^2_\mathrm{RS} - \frac{2 a_\mathrm{u}}{1 + a^2_\mathrm{u}} } \Bigg\} \left( \frac{t_\mathrm{on4}}{t_\mathrm{off}}\right) \approx  
    \begin{cases}
        &\ 1.28   \left( \frac{t_\mathrm{on4}}{t_\mathrm{off}}\right)  \hspace{1cm} \text{For $a_\mathrm{u} = 2$ and constant $L$} \\ 
        &\ 1.06   \left( \frac{t_\mathrm{on4}}{t_\mathrm{off}}\right)  \hspace{1cm} \text{For $a_\mathrm{u} = 5$ and constant $L$} 
    \end{cases}
\end{equation}

\begin{figure}
    \centering
    \includegraphics[scale=0.5]{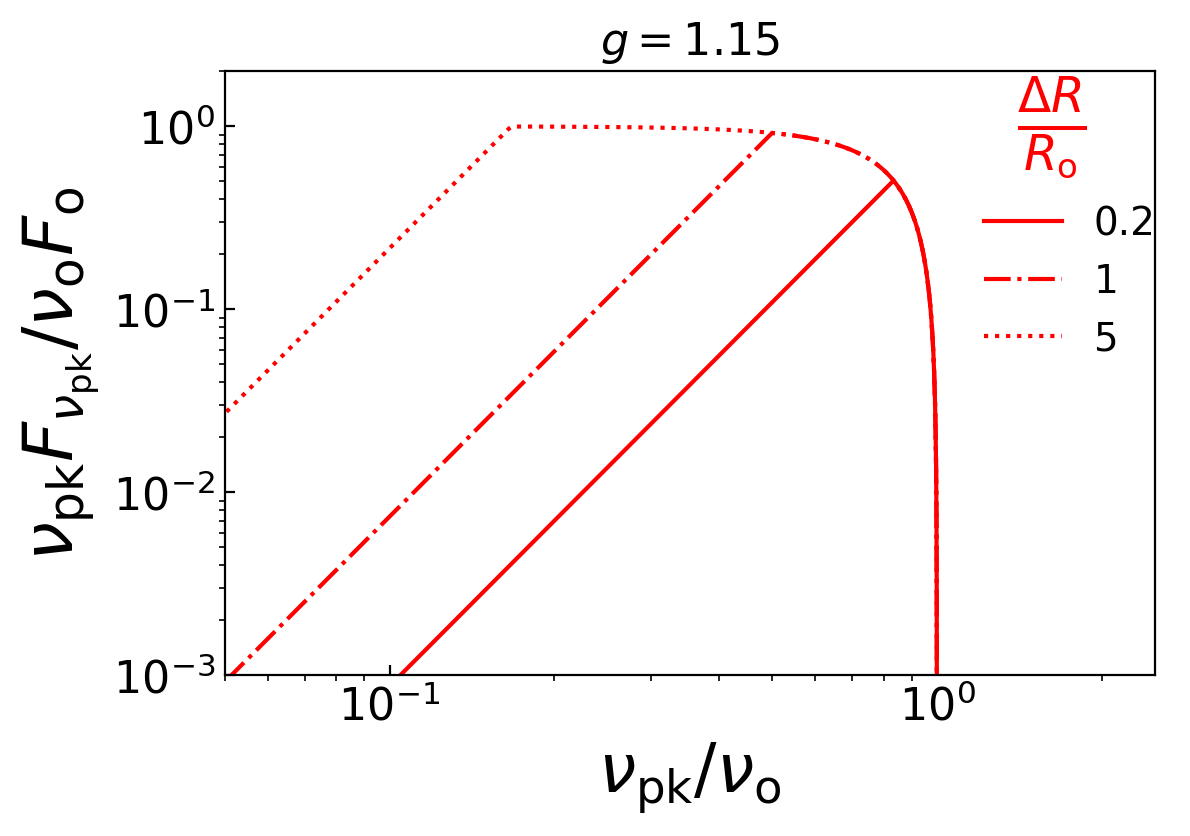}
    \caption{The behaviour of the reverse shocked instantaneous spectrum ($g_\mathrm{RS} = 1.15$) $\nu_\mathrm{pk} F_{\nu_\mathrm{pk}}$ vs $\nu_\mathrm{pk}$ for different values of the ratio $\Delta R/R_\mathrm{o}= (0.2,1,5)$.}
    \label{nupkFpk}
\end{figure}

The ratio $\Delta R/R_\mathrm{o}$ can be obtained analytically (within an error of 3 percent) as 
\begin{equation}
     \frac{\Delta R}{R_\mathrm{o}} \approx 
    \begin{cases}
        &\ \left(  \frac{\nu_\mathrm{pk,1/n}}{\nu_\mathrm{pk,max}}\right) \frac{1}{g^2_\mathrm{RS}} \left[ \left( 1 - \frac{1}{n} \right)^{-1/3} + (g^2_\mathrm{RS} - 1) \right] -1 \\  
    \end{cases}
\end{equation}
where $\nu_\mathrm{pk,max}$ is the frequency corresponding to the maximum value of $\nu_\mathrm{pk} F_{\nu_\mathrm{pk}} = (\nu_\mathrm{pk} F_{\nu_\mathrm{pk}})_\mathrm{max}$ and $\nu_\mathrm{pk,1/3}$ corresponds to the frequency when $(\nu_\mathrm{pk} F_{\nu_\mathrm{pk}})_\mathrm{1/n}$ is $1/n$ times the maximum value $(\nu_\mathrm{pk} F_{\nu_\mathrm{pk}})_\mathrm{max}$.

For $u_1 \gg 1$ and for a constant source power $L$, the $g_\mathrm{RS} = 1.15$ for a moderate proper speed contrast of $a_\mathrm{u} = 2$ and $g_\mathrm{RS} = 1.34$ for a high proper speed contrast of $a_\mathrm{u} = 5$. Assuming a representative value of $g_\mathrm{RS} = 1.15$ for $a_\mathrm{u} = 2$ and constant source power $L$, the ratio $\Delta R/R_\mathrm{o}$ and $t_\mathrm{on4}/t_\mathrm{off}$ for a sample of GRBs (data taken from Fig. 3 of \citealt{2023arXiv230800772Y}) has been shown in Table ~ \ref{promptGRBs} (Compare Fig. 3 of \citealt{2023arXiv230800772Y} and Fig. \ref{nupkFpk} from our internal shock model).

\begin{table}
    \centering
\caption{Estimation of $\Delta R/R_\mathrm{o}$ and $t_\mathrm{on4}/t_\mathrm{off}$ for a representative $g_\mathrm{RS} = 1.15$ and $a_\mathrm{u} = 2$. In the fourth and fifth column the numbers in the brackets corresponds to $\nu_\mathrm{pk,max}/\nu_\mathrm{pk,1/3}$. All data points are taken from Fig. 3 of \citealt{2023arXiv230800772Y}. }    
    \begin{tabular}{c|c|c|c|c} \hline  
      GRB   & $\nu_\mathrm{pk,max}/ \nu_\mathrm{pk,1/2}$  & $\nu_\mathrm{pk,max}/  \nu_\mathrm{pk,1/3}$ & $\Delta R/R_\mathrm{o}$ & $t_\mathrm{on4}/t_\mathrm{off}$ \\ \hline 
    GRB 140606B   & 0.45 & 0.38 & 1.66 (1.91) & 1.29 (1.49)  \\
    GRB 131011A   & 0.59 & 0.52 & 1.04 (1.09 ) & 0.81 (0.86)\\
    GRB 170607A   & 0.36 & $--$ & 2.36 & 1.85 \\ 
    GRB 151027A   & 0.65 &  0.57 & 0.84 (0.93) & 0.66 (0.72) \\
    GRB 150514A   & 0.60 & 0.53 & 0.99 (1.09) & 0.78 (0.85)\\
    GRB 120326A   & 0.70 &  0.57 & 0.71 (0.94) & 0.56 (0.74) \\
    GRB 190829A   & 0.50 &  0.43 & 1.40 (1.59) & 1.09 (1.24) \\  \hline 
    \end{tabular}
    \label{promptGRBs}
\end{table}


\bsp	
\label{lastpage}
\end{document}